\documentclass[preprint,aps,nofootinbib,superscriptaddress,floatfix]{revtex4-1}
\usepackage{graphics}
\usepackage{amsmath}
\usepackage{amssymb}
\usepackage{bm}
\usepackage{color}
\usepackage{dcolumn}
\usepackage{hyphenat}
\usepackage{hyperref}
\usepackage{url}
\usepackage{enumitem}
\usepackage{subfigure}

\allowdisplaybreaks

\def\be{\begin{equation}}
\def\ee{\end{equation}}
\def\ba{\begin{eqnarray}}
\def\ea{\end{eqnarray}}

\def\f{\frac}

\def\hub{{\mathcal H}}

\def\ie{{\frenchspacing\it i.e.}}

\newcommand{\CITE}[1]{[{\bf CITE}]}
\definecolor{darkviolet}{rgb}{0.58, 0.0, 0.83}

\usepackage{graphicx}

\usepackage[normalem]{ulem}

\begin{document}

\title{Phenomenology of Horndeski Gravity under Positivity Bounds}
\author{Dani de Boe}
\email{deboe@lorentz.leidenuniv.nl}
\affiliation{Institute Lorentz, Leiden University, PO Box 9506, Leiden 2300 RA, The Netherlands}
\author{Gen Ye}
\email{ye@lorentz.leidenuniv.nl}
\affiliation{Institute Lorentz, Leiden University, PO Box 9506, Leiden 2300 RA, The Netherlands}
\author{Fabrizio Renzi}
\affiliation{Institute Lorentz, Leiden University, PO Box 9506, Leiden 2300 RA, The Netherlands}
\author{In$\hat{\textrm{e}}$s S. Albuquerque}
\affiliation{Instituto de Astrof\'{i}sica e Ci\^{e}ncias do Espa\c{c}o, Faculdade de Ci\^{e}ncias da Universidade de Lisboa, Edif\'{i}cio C8, Campo Grande, P-1749-016 Lisbon, Portugal}
\author{Noemi Frusciante}
\affiliation{Dipartimento di Fisica ``E. Pancini", Universit\`a degli Studi  di Napoli  ``Federico II", Compl. Univ. di Monte S. Angelo, Edificio G, Via Cinthia, I-80126, Napoli, Italy}
\author{Alessandra Silvestri}
\affiliation{Institute Lorentz, Leiden University, PO Box 9506, Leiden 2300 RA, The Netherlands}

\begin{abstract}
A set of conditions that any effective field theory needs to satisfy in order to allow for the existence of a viable UV completion has recently gained attention in the cosmological context under the name of \emph{positivity bounds}. In this paper we revisit the derivation of such bounds for Horndeski gravity and translate them into a complete set of viability conditions in the  language of effective field theory of dark energy. We implement the latter into  \texttt{EFTCAMB} and explore the large scale structure phenomenology of Horndeski gravity under positivity bounds. We build a  statistically significant sample of viable Horndeski models, and derive the corresponding predictions for the background evolution, in terms of $w_{\rm DE}$, and the dynamics of linear perturbations, in terms of the phenomenological functions $\mu$ and $\Sigma$, associated to clustering and weak lensing, respectively. We find that the addition of positivity bounds to the traditional no-ghost and no-gradient conditions considerably tightens the theoretical constraints on all these functions. The most significant feature is a strengthening of  the correlation $\mu\simeq\Sigma$, and a related tight constraint on the luminal speed of gravitational waves $c^2_T\simeq1$. In anticipation of a more complete formulation of positivity conditions in cosmology, this work demonstrates the strong potential of such bounds in shaping the viable parameter space of scalar-tensor theories.

\end{abstract}

\maketitle

\section{Introduction}
\label{sec:introduction}
Effective field theories (EFTs) are a convenient model building tool that allow one to focus on the physics relevant within a given range of energy, while ignoring degrees of freedom at higher energies, thus making the problem manageable. They are ubiquitous in physics and have been key in achieving successful descriptions of many physical phenomena, starting from particle physics to superconductivity~\cite{Weinberg:1980wa,Wilson:1983xri,Penco:2020kvy,Weinberg:2021exr,Baumgart:2022yty}. In cosmology, EFTs have been applied to  inflation~\cite{Weinberg:2008hq,Cheung:2007st,Senatore:2010wk}, the study of large scale structure (LSS)~\cite{Baumann:2010tm,Carrasco:2012cv,Hertzberg:2012qn} and, more recently, dark energy (DE)~\cite{Bloomfield:2012ff, Gubitosi:2012hu,Piazza:2013coa,Frusciante:2019xia}. EFTs are defined by an effective action for the relevant degrees of freedom, built in terms of all operators that obey a predefined set of symmetry principles and following an expansion in some parameter. The latter is a small quantity representing the separation between the physics one is neglecting and the degrees of freedom considered, allowing to control the impact of the former on the latter.  Each operator in the EFT action, is then accompanied by a coefficient which is proportional to the expansion parameter, often referred to as Wilson coefficients.

The EFT approach thus introduces a number of  free coefficients in the action, which in a cosmological setting are even promoted to  free functions of time. It then becomes important to envisage ways of constraining this parameter space. 
In Minkowski spacetime, it is known that deep insight on the EFT parameter space can be made by studying the potential complete UV embedding of the EFT. While sometimes the complete theory is known, and working with an EFT is simply a matter of convenience, more often the UV completion is simply unknown. Either way, the simple assumption that the UV complete theory is consistent with the Wilsonian description, i.e. that it obeys unitarity, causality and Lorentz invariance, can be projected into positive bounds on the amplitudes  for 2-2 scattering at low energies. These result into a set of conditions that the Wilsonian coefficients of the EFT action need to satisfy and are often referred to as \emph{positivity bounds}~\cite{Adams:2006sv}, see also~\cite{Joyce:2014kja} for a pedagogical review. 

While originally formulated for massive fields on Minkowski spacetime, there has recently been growing interest in importing these bounds into cosmology~\cite{Nicolis:2009qm, Baumann:2015nta, Bellazzini:2015cra, Cheung:2016yqr, Cheung:2016wjt, deRham:2017imi, Bellazzini:2017fep,deRham:2017xox,deRham:2017avq,Bellazzini:2019xts, Melville:2019wyy, Ye:2019oxx,Tokuda:2020mlf, Kennedy:2020ehn,deRham:2020zyh,deRham:2021fpu,Traykova:2021hbr,Grall:2021xxm,Melville:2022ykg,Xu:2023lpq,Bellazzini:2019xts,Bellazzini:2023nqj}. Until today, breaking of Lorentz invariance and time translation symmetry induced by the cosmological evolution remains to be the biggest obstacle in deriving rigorous positivity bounds in cosmology~\cite{Baumann:2015nta,Ye:2019oxx,Grall:2021xxm}. Nevertheless, there have been some effort exploring the impact of positivity bounds on the parameters of scalar-tensor and massive gravity theories~\cite{Baumann:2015nta, Melville:2019wyy, Ye:2019oxx, Kennedy:2020ehn,deRham:2020zyh,deRham:2021fpu,Traykova:2021hbr,Grall:2021xxm,Melville:2022ykg,Xu:2023lpq}, relying on several assumptions that we will revisit. In a similar fashion, in this paper we focus on gauging the impact of positivity bounds on the cosmological phenomenology within the full class of Horndeski gravity. Given the obstacles mentioned above, and as we will discuss,  in practice we assume that the Minkowski positivity bounds can be straightforwardly upgraded to a time-dependent cosmological setup. While these might not fully represent the positivity bounds that shall eventually emerge from a rigorous formulation on an expanding background, they provide the ground for a  first exploration of the effects on the cosmological gravitational landscape. 

We focus on the effects of positivity bounds on the expected phenomenology of Horndeski theory at the level of background and LSS. Following previous works from some of the authors~\cite{Raveri:2017qvt,Peirone:2017ywi,Espejo:2018hxa,Frusciante:2018vht}, we use the EFT of DE (hereafter EFTofDE) formulation of Horndeski theory, but adopt a new and more parameterization-independent method based on Gaussian Process (GP), to sample the theory space under the usual stability (e.g. no-ghost, no-gradient, see e.g. \cite{Frusciante:2016xoj,DeFelice:2016ucp} for their formulation in EFT) conditions  plus the new conditions contributed by positivity. We then use \texttt{EFTCAMB}~\cite{Hu:2013twa,Raveri:2014cka}, a public patch to the Einstein-Boltzmann solver \texttt{CAMB}~\cite{Lewis:1999bs,camb}, to project the resulting viable theory space, into theoretical distribution for a set of functions that capture the background and LSS dynamics, namely: the equation of state of dark energy, $w_{\rm DE}$; the gravitational effects on matter, captured by the function $\mu$ that relates the Newtonian potential to matter density contrast; the gravitational effect on light, captured by the function $\Sigma$ relating the Weyl potential to the matter density contrast. We compare the resulting {\it theoretical priors} for $w_{\rm{DE}},\mu$ and $\Sigma$ to those obtained, always for Horndeski theory, in previous works that did not include positivity bounds, showing that the latter lead to narrower distribution functions and  to a  strengthening of the condition $\mu\simeq\Sigma$. In view of this, we look also at the the {\it gravitational slip} $\gamma$, which is related to the former functions via $\gamma=2\Sigma/\mu-1$, showing how positivity bounds narrowly constrain it to be $\sim 1$. Correspondingly, we find a narrow theoretical distribution for the speed of tensors, $c_T^2$. 

The paper is organized as follows. In Section~\ref{Sec:poshorndeski} we derive the positivity bounds for Horndeski theories of gravity, discussing all the assumptions in our analysis, as well as some caveats. In Section~\ref{sec:code_notation} we describe the reconstruction procedure to map the covariant positivty bounds obtained in Section~\ref{Sec:poshorndeski} to the corresponding EFTofDE form, which is then implemented into \texttt{EFTCAMB}. In Section~\ref{Sec:Phenomenology} we study the impact of positivity bounds on the background and LSS phenomenology of Horndeski theories. We then conclude in Section~\ref{Sec:conclusion}.
 
\section{Positivity Bounds for Horndeski}\label{Sec:poshorndeski}
Horndeski gravity, originally formulated in 1974~\cite{Horndeski:1974wa}, represents the general, Lorentz-invariant and local scalar-tensor theory propagating one massless graviton and one scalar degree of freedom with second order equations of motion; the latter field,  can be non-minimally coupled to gravity, have derivative self-interactions as well as kinetical mixing with gravity. As reviewed in~\cite{Horndeski:2024sjk}, in the past decade, this theory has gained a lot of attention in cosmology  because it encompasses  most alternative models of gravity and dark energy, which could be relevant for the dynamics of the Universe on large scales and will undergo scrutiny  with the advent of Stage IV Large Scale Structure surveys such as Euclid~\cite{euclid} and the Vera Rubin Observatory~\cite{lsst}. 

As it was shown in~\cite{Kobayashi:2011nu}, the original action of Horndeski gravity is dynamically equivalent to that of  Generalized Galileons, which derives from a covariantization of the Galileons introduced on flat space in~\cite{Nicolis:2008in}. In this work we will use the Generalized Galileon action written in the following convention
\begin{equation}\label{eq:Horndeskie}
S = \int d^4 x \sqrt{-g} \sum^5_{i=2} \mathcal{L}_i,
\end{equation}
where $g$ is the  determinant of the metric tensor and the Lagrangians $\mathcal{L}_i$ are defined by 
\begin{align}
\mathcal{L}_2&=\Lambda_2^4 G_2(\phi,X),\nonumber \\
\mathcal{L}_3&=\Lambda_2^4 G_3(\phi,X) [\Phi], \nonumber \\
\mathcal{L}_4&= M_{\rm pl}^2 G_4(\phi,X) R + \Lambda_2^4 G_{4,X}(\phi,X)([\Phi]^2 - [\Phi^2]),\nonumber \\
\mathcal{L}_5&= M_{\rm pl}^2 G_5(\phi,X) G_{\mu \nu}\Phi^{\mu \nu} - \frac{1}{6}\Lambda_2^4 G_{5,X}(\phi,X)([\Phi]^3 - 3[\Phi][\Phi^2] + 2[\Phi^3]), \nonumber
\end{align}
where $G_i$ are functions of the dimensionless scalar field $\phi/M_{\mathrm{pl}}$ and $X = -\frac{1}{2}\nabla^\mu \phi \nabla_\mu \phi /\Lambda_2^4$, with $\Lambda_2^2 = M_{\mathrm{pl}}H_0$ ($H_0$ being the Hubble constant). The subscript $,\phi$ and $,X$  indicate partial derivation with respect to $\phi/\Lambda_1$ and $X$, respectively.  We have also defined  $\Phi^{\mu}_{ \ \nu}\equiv \nabla^\mu \nabla_\nu \phi/\Lambda_3^3$, with $\Lambda_3^3 = M_{\mathrm{pl}} H_0^2$, while square brackets indicate the trace, e.g. $[\Phi^2] = \nabla^\mu \nabla_\nu \phi \nabla^\nu \nabla_\mu \phi/\Lambda_3^6$. 

While being fully covariant and providing a complete action to explore cosmology both at early and late times~\cite{DeFelice:2010nf,Kobayashi:2011nu}, action~\eqref{eq:Horndeskie} contains several non-renormalizable terms and, from a fundamental point of view, should be considered an effective, low energy description of a UV complete theory. Without worrying about the details of the latter, we can derive a set of conditions that the coefficients $G_i$ need to satisfy in order for the UV completion to be consistent with a standard Wilsonian field theory description~\cite{deRham:2017imi,Bellazzini:2017fep,deRham:2017xox,Bellazzini:2019xts}. In what follows, we revisit the derivation of such conditions, with the goal of being self-contained and also highlighting the different caveats that one encounters and still need to be fully resolved. We defer the most technical parts, including some long equations and the Feynmann diagrams, to Appendix~\ref{App:pos_bounds}. As originally advocated by~\cite{Adams:2006sv}, the requirement of a causal, unitary and local UV completion translates into a set of bounds, of which the simplest apply to  the tree-level scattering amplitude of two massive particles on a flat Minkowski background. The latter can be expanded  in terms of the center-of-mass energy, $s$, and the momentum transfer, $t$ and the bounds require  that the expansion coefficients be positive (as a direct consequence of the optical theorem in the forward limit). Hence the name \emph{positivity bounds}. We will now focus on action~\eqref{eq:Horndeskie}, identify the relevant scattering processes and derive the corresponding  scattering amplitudes.

We start expanding the metric around a flat Minkowski background, \ie \ $g_{\mu \nu} = \eta_{\mu \nu} + h_{\mu \nu}/M_{\mathrm{pl}}$ and the scalar field around its vacuum expectation value $\langle \phi \rangle$, \ie \ $\phi = \langle \phi \rangle + \varphi$. Since the vacuum is Poincar{\'e}-invariant we can choose $\langle \phi \rangle = 0$ without loss of generality. Correspondingly, the functions $G_i$ can be expanded as follows~\cite{Hohmann:2015kra}: 
\begin{equation}
G_i = \sum_{n,m=0}^\infty \frac{1}{n! m!} \Big(\frac{\partial^{n+m} G_i}{\partial^n (\phi/M_{\mathrm{pl}}) \partial^m X}\Big)_{\phi = 0, X = 0} \Big(\frac{\varphi}{M_{\mathrm{pl}}}\Big)^n Y^m, 
\end{equation}
where $Y = -\frac{1}{2\Lambda_2^4} \nabla_\mu \varphi \nabla^\mu \varphi$. Evaluation of quantities at the background spacetime will be indicated by overbars. 

The tree-level scattering amplitude for $hh \rightarrow hh$ and $h \varphi \rightarrow h\varphi$ vanish at order $\mathcal{O}(1/M_{\mathrm{pl}})$~\cite{Melville:2019wyy}, therefore we will focus on the $\varphi \varphi \rightarrow \varphi \varphi$ scattering. The relevant vertex terms for the latter, up to $O\left(1/M_{\mathrm{pl}}\right)$ and after neglecting several boundary terms, are: 
\begin{eqnarray}
\mathcal{L}_{\varphi \varphi} &=& \frac{1}{2 M_{\mathrm{pl}}^2}\Lambda_2^4 \bar{G}_{2,\phi \phi}\varphi^2 -\frac{1}{2} \bar{G}_{2,X} (\partial^\mu \varphi)(\partial_\mu \varphi) - \bar{G}_{3,\phi} (\partial^\mu \varphi)(\partial_\mu \varphi)-\frac{1}{2}m^2 \varphi^2\,, \nonumber \\
\mathcal{L}_{hh} &=& \frac{\bar{G}_4}{8} \partial^\alpha h \partial_\alpha h - \frac{\bar{G}_4}{4} (\partial_\lambda h^\mu_\nu)(\partial^\lambda h^\nu_\mu)\,, \nonumber \\
\mathcal{L}_{\varphi \varphi \varphi} &=& \frac{\bar{G}_{2,X\phi} + 2 \bar{G}_{3,\phi \phi}}{4M_{\mathrm{pl}}} \varphi^2 \partial_\mu \partial^\mu \varphi + \frac{\bar{G}_{3,X} + 3\bar{G}_{4,X\phi}}{3 \Lambda_3^3} \delta^{\mu \nu}_{\alpha \beta} \varphi \partial_\mu \partial^\alpha \varphi \partial_\nu \partial^\beta \varphi\,, \nonumber \\
\mathcal{L}_{h \varphi \varphi} &=& -\frac{\bar{G}_{4,X} - \bar{G}_{5,\phi}}{2\Lambda_3^3} \varphi \delta^{\mu \nu \rho}_{\alpha \beta \kappa} \partial^\alpha \partial_\mu \varphi \partial^\beta \partial_\nu h^\kappa_\rho + \frac{\bar{G}_{2,X} + 2\bar{G}_{3,\phi}}{2M_{\mathrm{pl}}} h_{\mu \nu}(\partial^\mu \varphi \partial^\nu \varphi - \frac{1}{2}\eta^{\mu \nu} \partial^\alpha \varphi \partial_\alpha \varphi)\nonumber\\
&-& \frac{\bar{G}_{4,\phi \phi}}{2M_{\mathrm{pl}}} \varphi^2 \delta^{\mu \nu}_{\alpha \beta} \partial^\alpha \partial_\mu h^\beta_\nu\,, \nonumber \\
\mathcal{L}_{\varphi \varphi \varphi \varphi}& = &\frac{3 \bar{G}_{4,XX} - 2\bar{G}_{5,\phi X}}{12 \Lambda_3^6} \varphi \delta^{\mu \nu \rho}_{\alpha \beta \gamma} \partial^\alpha \partial_\mu \varphi \partial^\beta \partial_\nu \varphi \partial^\gamma \partial_\rho \varphi + \frac{3\bar{G}_{2,XX} + 4 \bar{G}_{3,X\phi}}{24 M_{\mathrm{pl}} \Lambda_3^3} \partial_\mu \varphi \partial^\mu \varphi \partial_\nu \varphi \partial^\nu \varphi \nonumber\\
&+&\frac{\bar{G}_{3,X\phi}+ 3 \bar{G}_{4,X\phi \phi}}{6M_{\mathrm{pl}}\Lambda_3^3}\varphi^2 \delta^{\mu \nu}_{\alpha \beta}\partial^\alpha \partial_\mu \varphi \partial^\beta \partial_\nu \varphi, 
\label{eq:LagHorn}
\end{eqnarray}
where we have added the usual gauge-fixing term $\mathcal{L}_{\mathrm{GF}} = -\frac{\bar{G}_4}{2} \Big(\partial^\mu h_{\mu \nu} - \frac{1}{2} \partial_\nu h\Big)^2$ to the graviton propagator.

We report in Appendix~\ref{App:pos_bounds} the corresponding  Feynman diagrams and rules to calculate the amplitudes. 
For an elastic  scattering in the decoupling limit ($M_{\mathrm{pl}}\rightarrow \infty$ while keeping $\Lambda_3$ fixed) the resulting tree-level  amplitude, expanded in powers  $s$ and $t$, (and safely neglecting loop corrections up to order $\mathcal{O}(\Lambda_3^{-6})$~\cite{Tokuda:2020mlf}), reads~\cite{Melville:2019wyy}: 
\begin{equation}\label{eq:ScatA}
\mathcal{A}(s,t) = c_{ss} \frac{s^2}{\Lambda_2^4} + c_{sst} \frac{s^2 t}{\Lambda_3^6} + ... \,\,,
\end{equation}
 with 
 \begin{equation}\label{eq:posboundHorn1}
\frac{2}{3}c_{sst}=-\bar{G}_{4,XX} + \frac{2}{3} \bar{G}_{5,\phi X} + \frac{1}{2} \frac{(\bar{G}_{3,X} + 3\bar{G}_{4,X\phi})^2}{2\bar{G}_{3,\phi} + \bar{G}_{2,X}} 
- \frac{1}{\bar{G}_4} (\bar{G}_{4,X} - \bar{G}_{5,\phi})^2 \,, 
\end{equation}
and
\begin{eqnarray}
&c_{ss}=\frac{1}{2} \bar{G}_{2,XX} - 2 \bar{G}_{4,X\phi \phi} - \frac{1}{2} \bar{G}_{2,\phi \phi} \frac{(\bar{G}_{3,X} + 3 \bar{G}_{4,\phi X})^2}{(2\bar{G}_{3,\phi} + \bar{G}_{2,X})^2} 
+ \frac{(\bar{G}_{2,X\phi} + 2\bar{G}_{3,\phi \phi})(\bar{G}_{3,X} + 3\bar{G}_{4,X \phi})}{2 \bar{G}_{3,\phi}+ \bar{G}_{2,X}}\nonumber\\
&\nonumber\\
&+ \frac{1}{\bar{G}_4} (\bar{G}_{5,\phi} - \bar{G}_{4,X})(2\bar{G}_{4,\phi \phi} - 2 \bar{G}_{3,\phi} - \bar{G}_{2,X})  \,.\label{eq:posboundHorn2}
\end{eqnarray}
 The above relations do not assume the
 canonical normalization $2\bar{G}_{3,\phi}+\bar{G}_{2,X} = 1$ but we let $2\bar{G}_{3,\phi} + \bar{G}_{2,X}>0$ be general. We also safely assume $\bar{G_4}\neq0$.
The result is consistent with Ref.~\cite{Tokuda:2020mlf} upon imposing the canonical normalization $2\bar{G}_{3,\phi}+\bar{G}_{2,X}=1$, $\bar{G}_4 = 1/2$, but has some mismatches with Ref.~\cite{Melville:2019wyy}, which is also noticed by Ref.~\cite{Tokuda:2020mlf}. 
 
In order to ensure the existence of a sensible UV completion, the coefficients in Eq.~\eqref{eq:ScatA} need to obey \cite{deRham:2017avq}
\begin{equation}
c_{ss}\geq 0 \quad c_{sst}\geq -c_{ss}\frac{3\Lambda_3^4}{2\Lambda_2^4}\,,
\end{equation}
which, under the assumption that $\Lambda_2 \gg \Lambda_3$ ~\cite{Melville:2019wyy} become simply $c_{ss} \geq 0$ and $ c_{sst} \geq 0$.

\subsection{Assumptions and caveats.}
\label{subsec:assumptions}
A Minkowski background is considered to derive the positivity bounds, therefore one may wonder about their applicability and validity in a cosmological setting. 
Deriving positivity bounds on a cosmological background is non-trivial because both time translation and boost symmetries are broken due to the fact that the Universe is expanding. 
The broken time translation symmetry often leads to an ill-defined $S$-matrix theory~\cite{Grall:2021xxm} and therefore other methods will be needed. Additionally, scattering amplitudes involving gravitons on curved backgrounds are not well understood yet since they lead  to a massless $t$-pole. One first attempt to compute   the positivity bounds on a cosmological background has been done in the context of Horndeski theory by making some assumptions: time translations are only very weakly broken (\ie \ one assumes that the EFT functions vary slowly compared to the Hubble rate $H$); boosts are completely broken; the scalar fields do decouple from the gravitons (such that the massless $t$-pole issue can be circumvented) and it assumes sub-horizon scales ($k \gg aH$)~\cite{Grall:2021xxm}. In an expanding Universe the dynamics of cosmological evolution are characterized by the Ricci curvature $R\sim H^2\sim \dot{H}$, thus a naive order of magnitude estimation of leading cosmological correction to the positivity bounds is $\mathcal{O}(H^2/\Lambda^2,\dot{H}/\Lambda^2)$ with $\Lambda$ being the EFT cutoff. Such a correction has been explicitly observed in the case of shift-symmetric beyond-Horndeski theory with $c_T=1$~\cite{Ye:2019oxx}. Because of the aforementioned complications and the fact that in EFTofDE $H/\Lambda\ll1$, we assume that the positivity bounds, Eqs. \eqref{eq:posboundHorn1} and \eqref{eq:posboundHorn2}, still hold on a cosmological background. In the computation we will not take into account interactions between the scalar field and other matter species, however this can be done and will only lead to additional positivity bounds~\cite{deRham:2021fpu}. Under the assumption that matter interacts with gravitons only (thus neglecting self-interactions of matter) through a universal minimal coupling $h_{\mu \nu}T^{\mu \nu}/(2M_{\mathrm{pl}})$, it has been shown that this implies a (super-)luminal gravitional wave speed if there exists a spin-0 and/or spin-1 matter field that interacts in this fashion~\cite{deRham:2021fpu}. Ref.~\cite{Traykova:2021hbr} further suggests that a healthy Minkowski limit might be needed for positivity bounds to be valid in a theory. For Horndeski theories, this translates to the no-ghost, no-gradient and no tachyon conditions in the Minkowski limit, i.e. $\bar{G}_{2,X}+2\bar{G}_{3,\phi}>0$ and $\bar{G}_{2,\phi \phi} \leq 0$. We do not take into consideration these additional conditions in this paper. Including them will further tighten the constraints.

\section{Positivity Bounds For EFT of DE} \label{sec:code_notation}
The EFTofDE formulation~\cite{Gubitosi:2012hu,Bloomfield:2012ff} offers an ideal ground to perform analysis of the  theoretical viability of models~\cite{Raveri:2014cka, Frusciante:2016xoj}. Being formulated in terms of a unifying action with a handful of free functions of time, it offers also the opportunity of sweeping the gravitational landscape in an efficient way,  generating large ensembles of statistically independent models~\cite{Raveri:2017qvt}. Specifically, the linear dynamics of cosmological perturbations in Horndeski gravity can be fully described by the EFT action:
\begin{eqnarray}
\mathcal{S} &=& \int d^4x \sqrt{-g}  \bigg\{ \frac{m_0^2}{2} (1+\Omega(\tau))R + \Lambda(\tau) - c(\tau)\,a^2\delta g^{00} + \frac{m_0^2H_0^2\gamma_1(\tau)}{2} \left(a^2 \delta g^{00} \right)^2
 \nonumber \\
 &-&\frac{m_0^2H_0 \gamma_2(\tau)}{2} \, a^2\delta g^{00}\,\delta {K}{^\mu_\mu} -  \frac{m_0^2\gamma_3(\tau)}{2}\left[\left( \delta {K}{^\mu_\mu}\right)^2 - \delta {K}{^\mu_\nu}\,\delta {K}{^\nu_\mu}-\frac{ a^2}{2} \delta g^{00}\,\delta \mathcal{R}\right] \bigg\}+S_{m} [g_{\mu \nu}, \chi_m ],\nonumber\\
\label{EFT_action}
\end{eqnarray}
For ease of comparison and implementation in the code, we will use the \texttt{EFTCAMB} notation in this section, thus the relabelling of the Planck mass $M_{\rm pl} \to m_0$ in action \eqref{EFT_action}. The EFT formulation~\eqref{EFT_action} provides a unified way of solving background and linear perturbation evolution in Horndeski gravity numerically and has been implemented in \texttt{EFTCAMB}. We will refer to the free functions of time  $\{\Omega,\Lambda,c,\gamma_1,\gamma_2,\gamma_3\}$, multiplying the different operators in action~\eqref{EFT_action}, as the \emph{EFT functions}. The former three impact both the background and linear dynamics, while the $\gamma's$ affect only perturbations. The Friedmann equations can be used  to algebraically solve for $c$ and to formulate a differential equation for the Hubble parameter (as described in detail in Section~\ref{sec:methodology}), once $\Omega$ and $\Lambda$ have been specified. Consequently, the  full phenomenology of Horndeski gravity on large cosmological scales can be reproduced by choosing five free functions of time: $\{\Omega, \Lambda, \gamma_1,\gamma_2,\gamma_3\}$. 

To study the phenomenology of Horndeski models under positivity bounds with \texttt{EFTCAMB}, one needs to map the covariant positivity bounds Eqs.~\eqref{eq:posboundHorn1}-\eqref{eq:posboundHorn2} to the corresponding bounds in the EFT formulation. While any specific Horndeski model can be mapped into action~\eqref{EFT_action} with a unique set of the EFT functions~\cite{Bloomfield:2013efa, Gleyzes:2013ooa,Frusciante:2016xoj}, the inverse, i.e. to express the Horndeski $G_i$'s in terms of the EFT functions, is however much more involved and generally non-unique. Ref.~\cite{Kennedy:2017sof} reports a reconstruction procedure which, given a choice of EFT functions, provides a family of Horndeski models matching the corresponding EFT action. The reconstruction consists in writing the $G_i(\phi,X)$ as an expansion in powers of the kinetic term $G_i(\phi,X)=\sum_{n=0}^N g_{i,n}(\phi)X^n+\Delta G_i$ where $N=4-i$, with coefficients $g_{i,n}(\phi)$ which needs to be determined in terms of the EFT functions. This can be achieved by expanding the resulting Horndeski action, Eq.~\eqref{eq:Horndeskie}, with the above choices of $G_i$ up to second order in perturbations in the unitary gauge and imposing that it recovers the EFT action~\eqref{EFT_action}. Full expressions of $g_{i,n}$ are given in Appendix \ref{App:reconstruction}. Because $G_5$ does not contribute any new EFT operators to action~\eqref{EFT_action}, there is no corresponding deterministic part, i.e. $G_5(\phi,X)=\Delta G_5$. The $\Delta G_i$'s are higher order in $X$ and span a family of covariant theories that correspond to the same background and linear dynamics. In reality the major obstacle one is faced with is that linear dynamics only depend on quadratic order perturbed action while tree level positivity bounds require at least three point vertices, thus knowing only background and linear evolution (which is all one has in \texttt{EFTCAMB}) is not enough to specify a unique covariant theory to apply the tree level positivity bounds. \footnote{An alternative way around would be to study positivity bounds directly from the EFT action, e.g. in Ref.\cite{Ye:2019oxx}.}. In order to apply the bounds we have to introduce additional assumptions to pick out a particular theory from the family. To this end, we assume $\phi = m_0^{2}\,t$ with $t$ being the cosmic time (thanks to the invariance of the action under field redefinition) and $\Delta G_i=0$ (which also effectively sets $G_5\equiv0$). A possible caveat here is that while the last assumption on the $G_i$'s does not affect the linear dynamics in \texttt{EFTCAMB}, its impact on the positivity bound side is still unclear and requires further study, which we leave for future works.

With the aforementioned reconstruction procedure and assumptions, we are now able to cast Eqs.~\eqref{eq:posboundHorn1} and \eqref{eq:posboundHorn2} into bounds in terms of the EFT functions:
\begin{align}\label{eq:posbound1}
\frac{2}{3}c_{sst}=&\frac{1}{8(1+\Omega)m_0^6D} \left\{(1+\Omega)(4a^2H_0 \gamma_2^2
+8a\mathcal{H}^2 \gamma_3 \gamma_3^\prime + a^2 \mathcal{H}^2 \gamma_3^{\prime\,2})
+ 4(\mathcal{H}^2+2\dot{\mathcal{H}}) \gamma_3^3\right.\nonumber \\
&\left. +4a\mathcal{H}H_0 \gamma_2\left[(1+\Omega)(4\gamma_3+a\gamma_{3}^\prime)
+3\gamma_3^2\right] + 8\gamma_3^2\left[-\frac{ca^2}{m_0^2}+\mathcal{H}^2 (2(1+\Omega)+a\gamma_3^\prime)\right]\right\}\geq 0\,,
\end{align}
and
\begin{align}\label{eq:posbound2}
c_{ss}=&\frac{2\gamma_3 \Big[-D + a(\dot{\mathcal{H}} \Omega^\prime + a\mathcal{H}^2 \Omega^{\prime \prime})\Big]}{2a^2 m_0^6 (1+\Omega)} \nonumber \\ &+\frac{4a^2 H_0^2 \gamma_1 - aH_0\mathcal{H}(3\gamma_2 - a\gamma_2^\prime) + \Dot{\mathcal{H}} (4 \gamma_3 + a \gamma_3^{\prime}) - \mathcal{H}^2 (10\gamma_3 - a (\gamma_3^{\prime} + a \gamma_3^{\prime \prime}))}{2a^2 m_0^6} \nonumber \\
&+ \frac{2aH_0 \gamma_2 + \mathcal{H}(4\gamma_3 + a\gamma_3^\prime)}{2a^2 m_0^6 D}\Bigg[  3aH_0 (\gamma_2 \dot{\mathcal{H}} - \mathcal{H}^2(\gamma_2 - a\gamma_2^\prime)) - 2\mathcal{H}\dot{\mathcal{H}}(\gamma_3 - 3a\gamma_3^\prime)+ \nonumber \\
& 2\gamma_3 \ddot{\mathcal{H}} - \mathcal{H}^3(2\gamma_3 + a(\gamma_3^\prime - 2a\gamma_3^{\prime \prime}))-\frac{2\dot{c}a^2}{m_0^2}\Bigg]\nonumber \\
&-\frac{\left[2aH_0 \gamma_2 + \mathcal{H}(4\gamma_3 + a\gamma_3^\prime)\right]^2}{16a^2m_0^6D^2}\Bigg[ \frac{2a^2}{m_0^2}( \ddot{\Lambda}-\mathcal{H}\dot{\Lambda})- \mathcal{H}(a^2 H_0 \dot{\mathcal{H}}(\gamma_2^\prime-3a\gamma_2^{\prime \prime}) - \ddot{\mathcal{H}}(2\gamma_3 - 5a\gamma_3^\prime - 4a^2 \gamma_3^{\prime \prime}))- \nonumber \\
& 2\gamma_3 (2\dot{\mathcal{H}}^2 + \dddot{\mathcal{H}})
+a(aH_0 \gamma_2^\prime \ddot{\mathcal{H}} - \dot{\mathcal{H}}^2 (5\gamma_3^\prime + 3a\gamma_3^{\prime \prime}) - \gamma_3^\prime \dddot{\mathcal{H}}) + 2\mathcal{H}^2 \dot{\mathcal{H}}(18\gamma_3 - a^2(4\gamma_3^{\prime \prime}+3a\gamma_3^{\prime \prime \prime})) + \nonumber \\
&a^4 H_0 \mathcal{H}^3 \gamma_2^{\prime \prime \prime}-\mathcal{H}^4(24\gamma_3 - 12a\gamma_3^\prime+2a^3 \gamma_3^{\prime \prime \prime}+a^4 \gamma_3^{\prime \prime \prime \prime})\Bigg]  \geq 0. \,
\end{align}
where we have defined 
\begin{equation}
    D\equiv 2\frac{ca^2}{m_0^2}-3a\mathcal{H}H_0 \gamma_2 - (\mathcal{H}^2 +2\dot{\mathcal{H}})\gamma_3 - 2a \mathcal{H}^2 \gamma_3^\prime\,,
\end{equation}
with $\mathcal{H}=aH$ the Hubble parameter in conformal time, a prime indicating derivation with respect to the scale factor, and a dot derivation with respect to conformal time. Eqs.~\eqref{eq:posbound1} and~\eqref{eq:posbound2} are the final results of this section and are the actual positivity bounds we implemented into \texttt{EFTCAMB}. They are presented in the notation of the code for convenience of implementation, see Appendix~\ref{App:reconstruction} for some useful conversions between different notations in the literature. 
\section{Horndeski theory's Phenomenology under Positivity Bounds}
\label{Sec:Phenomenology}

Recent works~\cite{Melville:2019wyy,Kennedy:2020ehn,deRham:2021fpu,Traykova:2021hbr} have explored the impact of positivity bounds  on the parameter space of some specific cosmological models or class of modified gravity (i.e. shift-symmetric Horndeski theory). On the other hand, the phenomenology of Horndeski theory under ghost and gradient stability conditions has been investigated by some of the authors in~\cite{Raveri:2017qvt,Peirone:2017ywi,Espejo:2018hxa}, and including the tachyonic stability condition in~\cite{Frusciante:2018vht}. We refer the reader to the review paper~\cite{Frusciante:2019xia} for further discussion on the stability conditions. 
Here we will explore the space of theories that satisfy the positivity bounds, on top of complying with the traditional physical  stability requirements of absence of  ghosts and gradient instabilities, in the full class of Horndeski gravity. After sampling the theory space and evolving the cosmology of the viable models, we project the latter into the equation of state of dark energy, $w_{\rm DE}$, and the  two phenomenological functions  that describe the effects of gravity on matter and light at the level of  LSS, respectively $\mu$ and $\Sigma$. 

We adopt the following definition of the effective equation of state parameter of dark energy
\begin{align}
w_{\rm DE} \equiv \frac{P_{\rm DE}}{\rho_{\rm DE}} = \frac{-2\dot{\hub} -\hub^2 - P_m a^2/m_0^2}{ 3\hub^2 - \rho_m a^2/m_0^2 } \ ,
\label{eq:wDE}
\end{align}
where $\rho_m$ and $P_m$ are the combined energy density and the pressure of all matter species. A related quantity is the dark energy density parameter $\Omega_X$, defined via
\begin{equation}
    3m_0^2 H^2(a) = \rho_m(a) + 3H_0^2m_0^2\Omega_X(a).
    \label{eq:omega_X}
\end{equation}
As it is clear from its definition, $\Omega_X$ generally represents the contribution of any term other than  matter (here meant in the broad sense of relativistic and non-relativistic species). It  has the advantage of avoiding the singularities that arise in $w_{\rm DE}$ whenever the effective DE density crosses zero, as can happen in modified gravity models. 

The LSS functions are introduced in the Newtonian gauge 
\begin{equation}
    ds^2=-(1+2\Psi)dt^2+a^2(1-2\Phi)|d\vec{x}|^2
\end{equation}
through two equations that parametrize the relation between the Fourier transforms of the matter density contrast and the scalar perturbations to the metric, generalizing the Poisson equation as follows
\begin{align}\label{eq:mu_Sigma_def}
k^2 \Psi &= -4\pi G \mu(a,k) a^2 \rho \Delta, \nonumber \\
k^2 (\Psi + \Phi) &= -8\pi G \Sigma(a,k) a^2 \rho \Delta\,,
\end{align}
where $\Delta$ is the gauge invariant matter density contrast in the frame comoving with total matter, and we are focusing on late times, safely neglecting shear from neutrinos\footnote{It is possible to extend the definitions of $\mu$ and $\Sigma$ in order to account for shear from matter, so that the system of equations is valid also at early times~\cite{Zucca:2019xhg}.}. Along with the Euler and continuity equation for matter, Eqs.~\eqref{eq:mu_Sigma_def} form a complete set to evolve the dynamics of LSS~\cite{Pogosian:2010tj}. 
Phenomenologically, $\mu$ parametrizes modifications to the effective gravitational coupling relevant to matter clustering, and is best probed by galaxy clustering and redshift space distortions; $\Sigma$ parametrizes modifications to the Weyl potential and is best probed by measurements of the weak lensing of distant galaxies and CMB, as well as measurements of galaxy number counts through the so-called magnification bias~\cite{Amendola:2007rr,Jain:2007yk,Bertschinger:2008zb,Song:2010fg}; see~\cite{Planck:2018vyg,Pogosian:2021mcs,Raveri:2021dbu,DES:2022ccp} for some of the latest constraints.  $\mu$ and $\Sigma$  are equal to one in $\Lambda$CDM, but generally would be functions of time and the Fourier wavenumber $k$ in models beyond $\Lambda$CDM.

We will consider also the alternative function $\gamma$ commonly referred to as the \emph{gravitational slip}~\cite{Caldwell:2007cw,Daniel:2008et}, which parametrizes the ratio between the lensing and clustering potentials:
\begin{equation}
    \label{eq:grav_slip}
    \gamma\equiv\frac{\Phi}{\Psi}=\frac{2\Sigma}{\mu}-1\,.
\end{equation}
The gravitational slip is intimately related to  breaking of the equivalence principle and modifications of the propagation of tensors, and is therefore considered a “smoking gun” of modified gravity~\cite{Saltas:2014dha,Sawicki:2016klv}.
\subsection{Methodology}
\label{sec:methodology}
We are interested in studying how the positivity bounds~\eqref{eq:posbound1}-\eqref{eq:posbound2} impact the cosmological phenomenology of Horndeski gravity. To this extent,  we generate a large sample of random models respecting physical stability, i.e. ghost and gradient stability conditions~\cite{Frusciante:2016xoj}, as well as positivity bounds, calculate their cosmology and project it on the equation of state of dark energy $w_{\rm DE}$ and the LSS functions $\mu$, $\Sigma$ and $\gamma$. We compare the outcome with the same results obtained from  the sample of models obeying only physical stability. With large enough numbers of sampled models, any difference  in the distributions of  $\{w_{\rm DE}, \mu, \Sigma,
\gamma\}$ between the two samples encodes information about the impact of positivity bounds on the viable parameter space of Horndeski theory. 

We perform our model sampling by varying $\{\Omega,\Lambda, \gamma_1,\gamma_2,\gamma_3\}$, reconstructing the corresponding background dynamics and then evolving linear cosmological perturbations. All this is done within \texttt{EFTCAMB}. For a given choice of $\Omega(a), \Lambda(a)$ we combine the two 
 Friedmann equations to form a second order differential equation for the conformal Hubble parameter $\hub$. Following~\cite{Raveri:2017qvt}, and introducing $y \equiv \hub^2$, we have 
\begin{equation}
\left(1+\Omega+\f{1}{2}a\Omega'\right)\f{d y}{d \ln a} +\left(1+\Omega+2a\Omega' +a^2\Omega'' \right)y +\left( \frac{P_ma^2}{m_0^2} + \frac{\Lambda a^2}{m_0^2}\right) = 0\,,
\label{eq:hubble}
\end{equation}
where, like before, the prime indicates differentiation with respect to the scale factor. After solving this equation for the background, \texttt{EFTCAMB} 
computes and outputs both the background and 
the perturbation transfer functions, namely $\Delta(a,k), \Psi(a,k)$ and $\Phi(a,k)$, from which we calculate $w_{\rm DE}(a),~\mu(a,k),~\Sigma(a,k)$ and $\gamma(a,k)$ according to Eqs.~\eqref{eq:wDE},~\eqref{eq:mu_Sigma_def} and~\eqref{eq:grav_slip}.
 
We sample $\{\Omega,\Lambda,\gamma_i\}_{i=1,2,3}$ as functions of $x=\log_{10}a$  
adopting a GP approach where we draw each function from a Gaussian distribution of functions $f(x)\sim GP(\bar{f}, K)$ specified by the mean $\bar{f}(x)$ and kernel $K(x_1,x_2)$. Without prior knowledge on the theory, we adopt the general assumption that correlation between two time points, $x_1$ and $x_2$, decays as their distance in time $\Delta x=|x_1-x_2|$ increases, with the expectation that the models sampled are non-oscillatory. This leads us to choose the standard square exponential kernel
\begin{equation}\label{eq:gp_kernel}
    K(x_1,x_2) = \sigma^2\exp\left(-\frac{|x_1-x_2|^2}{2l^2}\right)
\end{equation}
for each of the functions sampled. For the mean functions, we use the GR/$\Lambda$CDM  limit of the EFT functions, specifically $\bar{\Omega}(x)=\bar{\gamma_1}(x)=\bar{\gamma_2}(x)=\bar{\gamma_3}(x)\equiv0$ and $\bar{\Lambda}(x)\equiv\Lambda_0$, where $\Lambda_0\equiv 3M_{\rm pl}^2(1-\Omega_m)H_0^2$ ($\Omega_m$ includes all matter species). Thus every EFT function introduces two parameters $\{\sigma,l\}$ to be sampled. In previous works~\cite{Raveri:2017qvt,Peirone:2017ywi,Espejo:2018hxa}, a Pad\'e  expansion for the EFT functions was used instead, with a total of $9$ parameters for each function. We opt for the GP method in order to maximize parameterization-independence while at the same time speeding up the convergence of the  procedure, thanks to the significantly smaller number of  parameters to sample. We have checked that the results are consistent with those of~\cite{Raveri:2017qvt,Peirone:2017ywi,Espejo:2018hxa} in the case of physical stability only.

\begin{table}[]
    \centering
    \begin{tabular}{|c|c|}
         \hline
         Parameters&Prior\\
         \hline
         $\omega_{\mathrm{m}}$&$\mathcal{N}(0.14, 0.04)$\\
         \hline
         $H_0$&$\mathcal{N}(70, 20)$\\
         \hline
         $l_A$, $A\in\{\Omega,\Lambda,\gamma_i\}$&Uniform$[0,5]$\\
         \hline
         $\sigma_A$, $A\in\{\Lambda,\gamma_i\}$&Uniform$[0,5]$\\
         \hline
         $\sigma_{\Omega}$&Uniform$[0,1]$\\
         \hline
    \end{tabular}
    \caption{Priors of the MCMC parameters. $\sigma_{\Omega}$ has a tighter prior because the variation of $\Omega$ between last-scattering and today is known to be small \cite{Aghanim:2018eyx} and  a bound at unity helps to reduce the occurrence of unrealistic models with $1+\Omega<0$ in the sampling process.}
    \label{tab:mcmc_priors}
\end{table}
We use the Monte Carlo Markov Chain (MCMC) method to sample over the cosmological parameters $\{\omega_{\rm{m}}, H_0\}$ and 10 GP parameters $\{\sigma_A, l_A\}_{A=\{\Omega,\Lambda,\gamma_i\}}$, in total 12 free parameters. We use uninformative wide priors for all parameters, see Table.\ref{tab:mcmc_priors}. At each parameter point $\{\omega_{\rm{m}}, H_0, \sigma_A, l_A\}_{A=\{\Omega,\Lambda,\gamma_i\}}$ we sample 50 GP random realizations of the EFT functions to explore the theoretical function space as much as possible. In practice, samples drawn from the GP sampler are discrete function values on a dense redshift grid, which is then interpolated to produce a smooth function using quintic interpolation. High interpolation order is necessary because the positivity bounds, Eqs.~\eqref{eq:posbound1} and \eqref{eq:posbound2}, are sensitive to up to fourth derivative of the EFT functions. We use \texttt{Cobaya}~\cite{Torrado:2020dgo} to perform the MCMC sampling and a new \texttt{EFTCAMB} version with physical stability and positivity conditions implemented to calculate cosmology and perform stability checks. We adopt the Gelman-Rubin statistic $R-1<0.05$ as our convergence criteria. To sample the relevant theory space as much as possible, we also ensure a large number ($\gtrsim\mathcal{O}(10^4)$) of models are accepted in each MCMC run.

\begin{figure}[t]
    \centering
    \includegraphics[width=\linewidth]{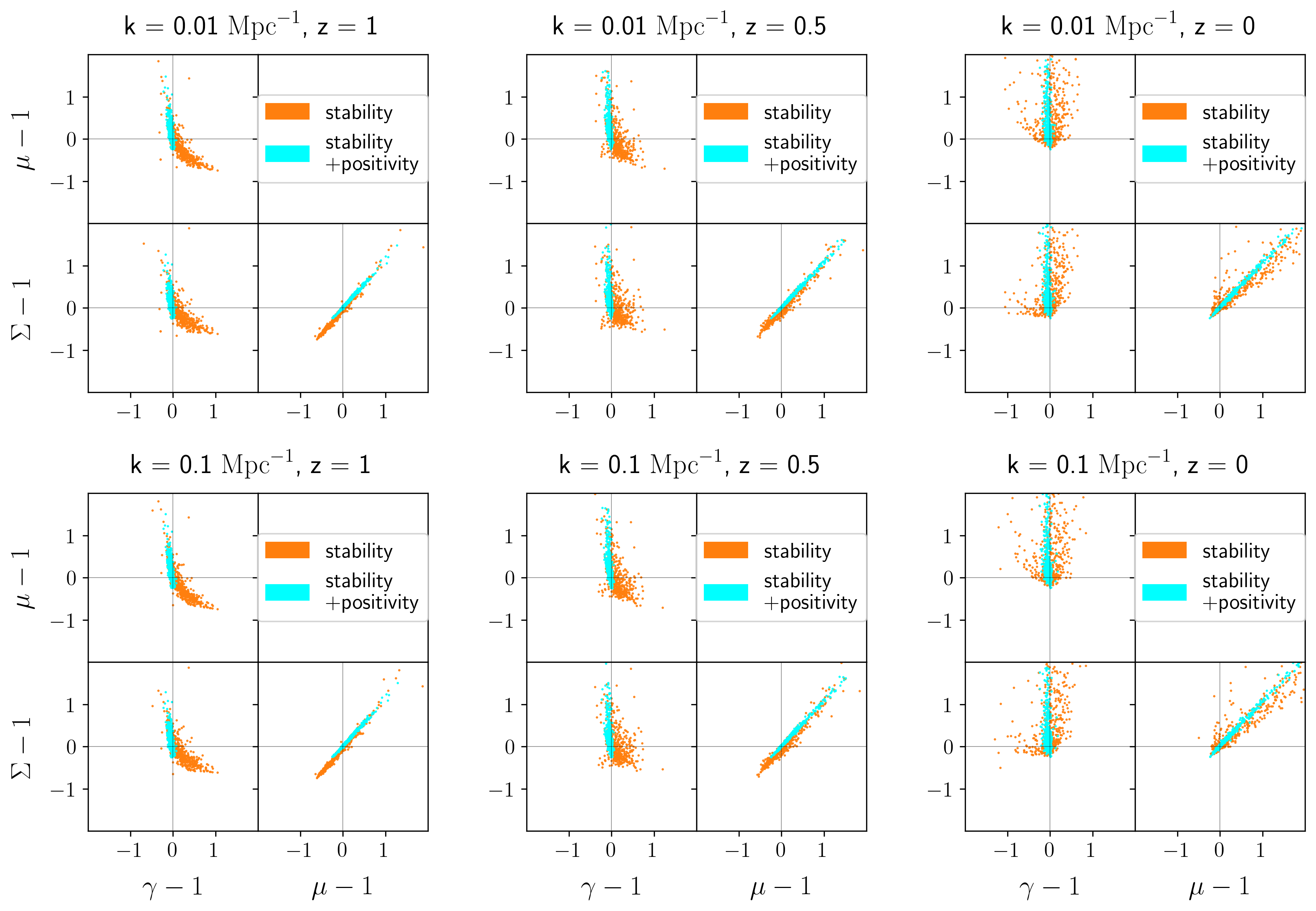}
    \caption{Distribution of the phenomenological LSS functions $\mu$ (clustering), $\Sigma$ (lensing) and $\gamma$ (gravitational slip) at $z\in\{0, 0.5, 1\}$ and $k \in \{0.01, 0.1\}\,\mathrm{Mpc}^{-1}$. Starting from the left column and moving right, we go from early to late times.  In each case, two set of sampled EFT models are displayed, one which satisfies the  ghost+gradient (stability) conditions only, shown in orange, and the other that satisfies stability+positivity bounds conditions, shown in cyan.}
    \label{fig:cloud}
\end{figure}
In the MCMC run, the physical stability and positivity conditions are treated as a pass/fail test of the sampled EFT model. A model is rejected if it violates any of the conditions. While this guarantees that the sample contains only models that are physically viable, it does not necessarily imply that out of  the infinite-dimensional theory space we have been exploring the area that provides a realistic description of our Universe. We therefore further require the viable models to obey some  observational and experimental priors as follows:
\begin{itemize}[leftmargin=*]
\item[B1.] The theory returns to GR at high redshift. This is achieved by turning off all EFT functions at $z>100$ using a smooth $\tanh$ step function.
\item[B2.] CMB and BBN constraints on the variation of the Planck mass from last-scattering to today \cite{Aghanim:2018eyx}. Combined with the above return-to-GR requirement, this amounts to $|\Omega(z=0)-1|<0.1$.
\item[B3.] Following the constraint on the speed of tensors from the GW170817/GRB170817A multi-messenger event~\cite{TheLIGOScientific:2017qsa,Monitor:2017mdv,Coulter:2017wya}, we set $\gamma_3(z=0)=0$ \cite{Baker:2017hug, Ezquiaga:2017ekz,Creminelli:2017sry}. 
\item[B4.] A very weak observational constraint from Type Ia Supernovae (SNIa) luminosity distance measurements based on the Pantheon data~\cite{Pan-STARRS1:2017jku}, with a significantly inflated covariance (by a factor of four)  to dissuade the chain from sampling extremely unrealistic cosmological backgrounds.
\end{itemize}

We emphasize that these additional  conditions have simply the aim of excluding models that would grossly violate known constraints, and do not correspond to a fit to data. Our focus remain on deriving theoretical priors.

\subsection{Results}
\label{sec:results}

\begin{figure}
    \centering
    \includegraphics[width=0.48\textwidth]{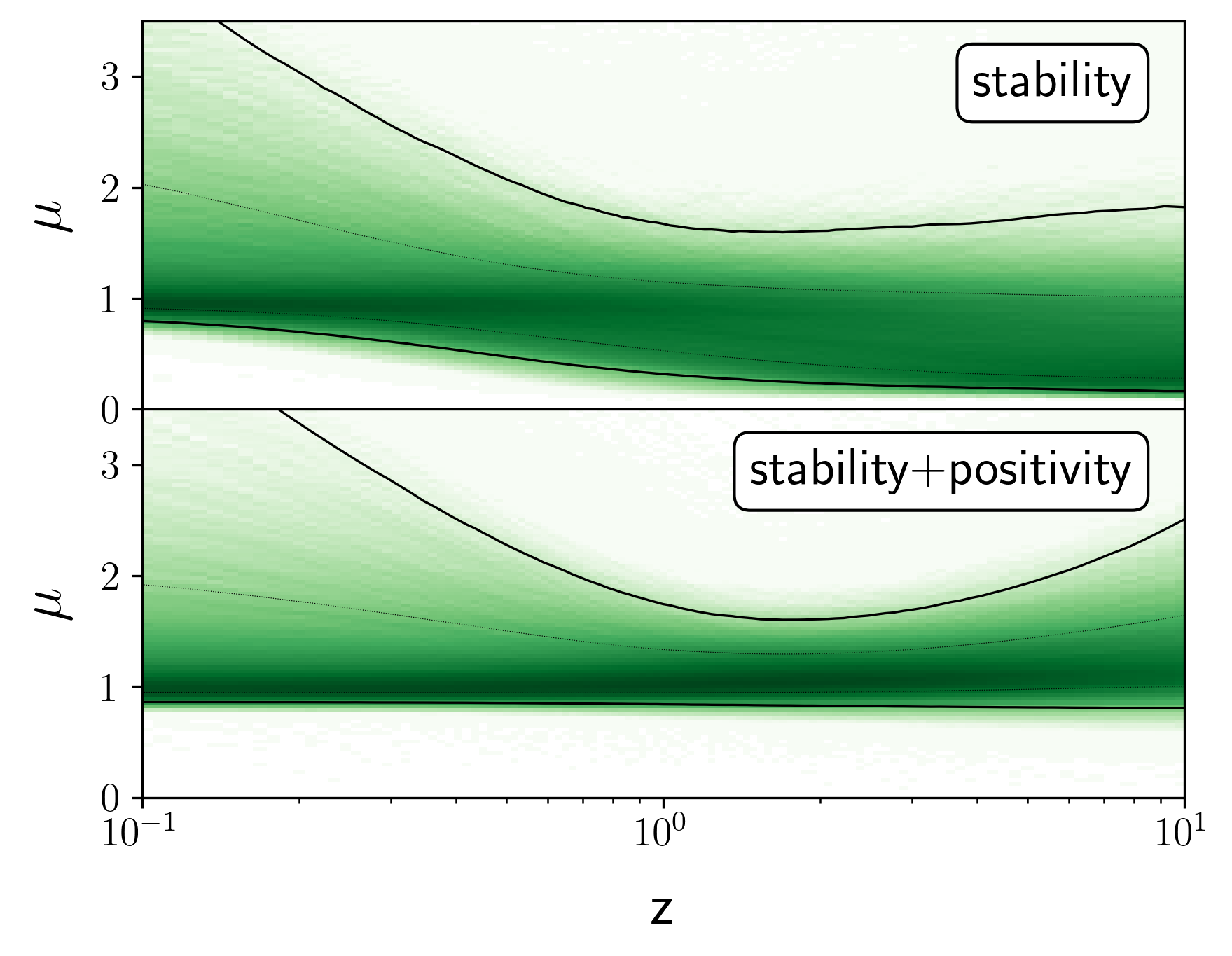}
    \includegraphics[width=0.48\textwidth]{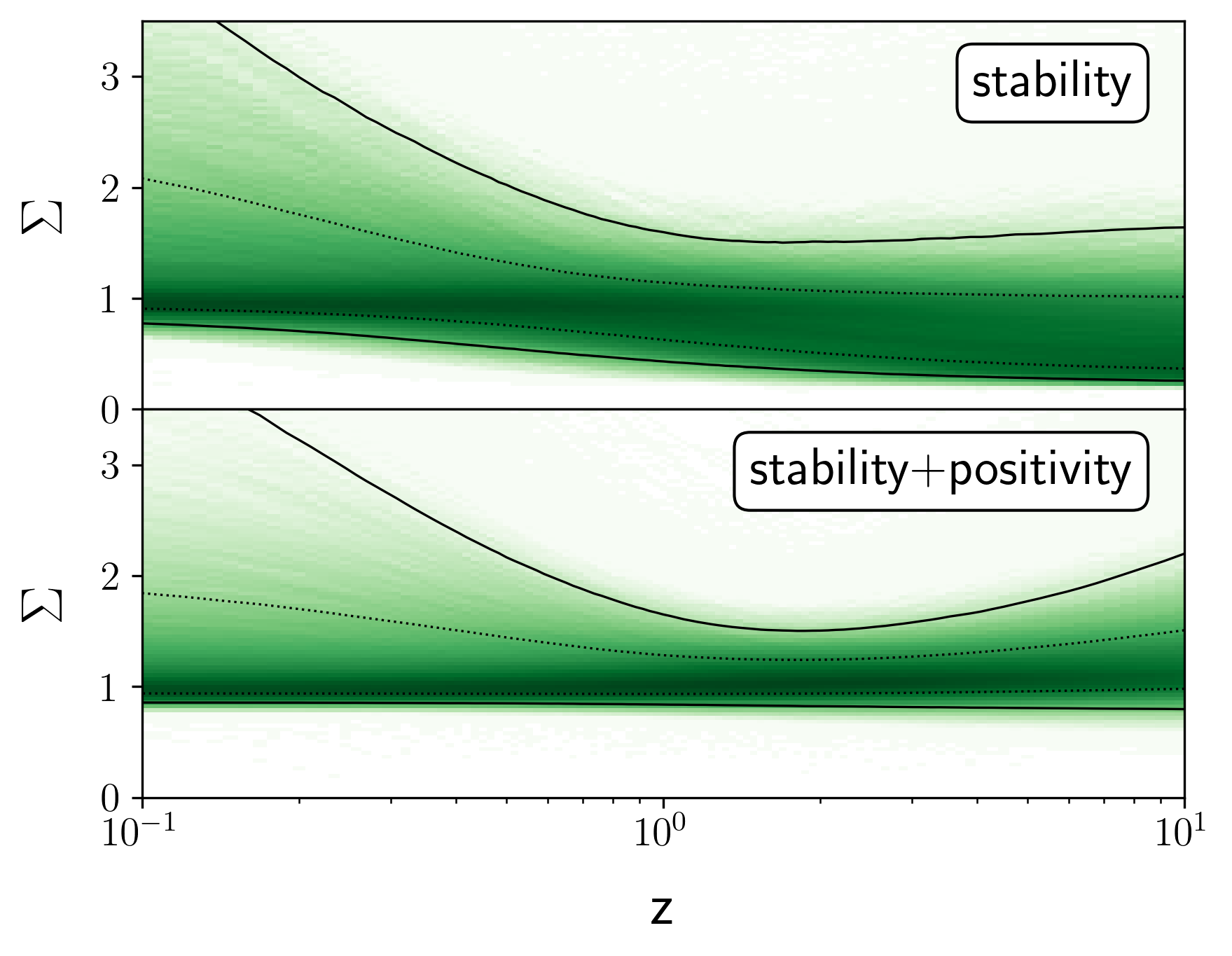}
    \caption{Theoretical distribution of the phenomenological functions $\mu(z)$ ({\it left panel}) and $\Sigma(z)$ ({\it right panel}) as functions of redshift at scale $k=0.1 \,{\rm Mpc}^{-1}$. Color shading represents the density of function values at each redshift with dark color indicating high density. Solid (Dotted) lines plot the region where 95\% (68\%) of the function values reside.}
    \label{fig:musgmgm}
\end{figure}

Our main results are summarized in Figure~\ref{fig:cloud}, which shows the distributions of viable Horndeski models, each corresponding to a point, in the planes of the LSS phenomenological functions $\mu$, $\Sigma$ and $\gamma$, at different redshifts, $z\in\{0, 0.5, 1\}$, and scales, $k\in \{0.01, 0.1\}\,\mathrm{Mpc}^{-1}$. We consider two cases: models that pass stability, along with the conditions B1-4; and models that on top of that, satisfy also the positivity bounds.  In both cases, the sampled points show a clear preference for the I and III quadrants in the $\mu-\Sigma$ plane, compatible with the $(\Sigma-1)(\mu-1)\ge0$ conjecture of Ref.~\cite{Pogosian:2016pwr}. Compared with the case of stability only, we can see that positivity bounds clearly add constraining power in all three variables, notably disfavoring models that lead to $\mu$ and $\Sigma$ considerably smaller than one. The biggest tightening concerns $\gamma$, the gravitational slip, which is now very close to $\gamma=1$, corresponding to $\Sigma\simeq\mu$. 
\begin{figure}
    \centering

    \includegraphics[width=0.48\textwidth]{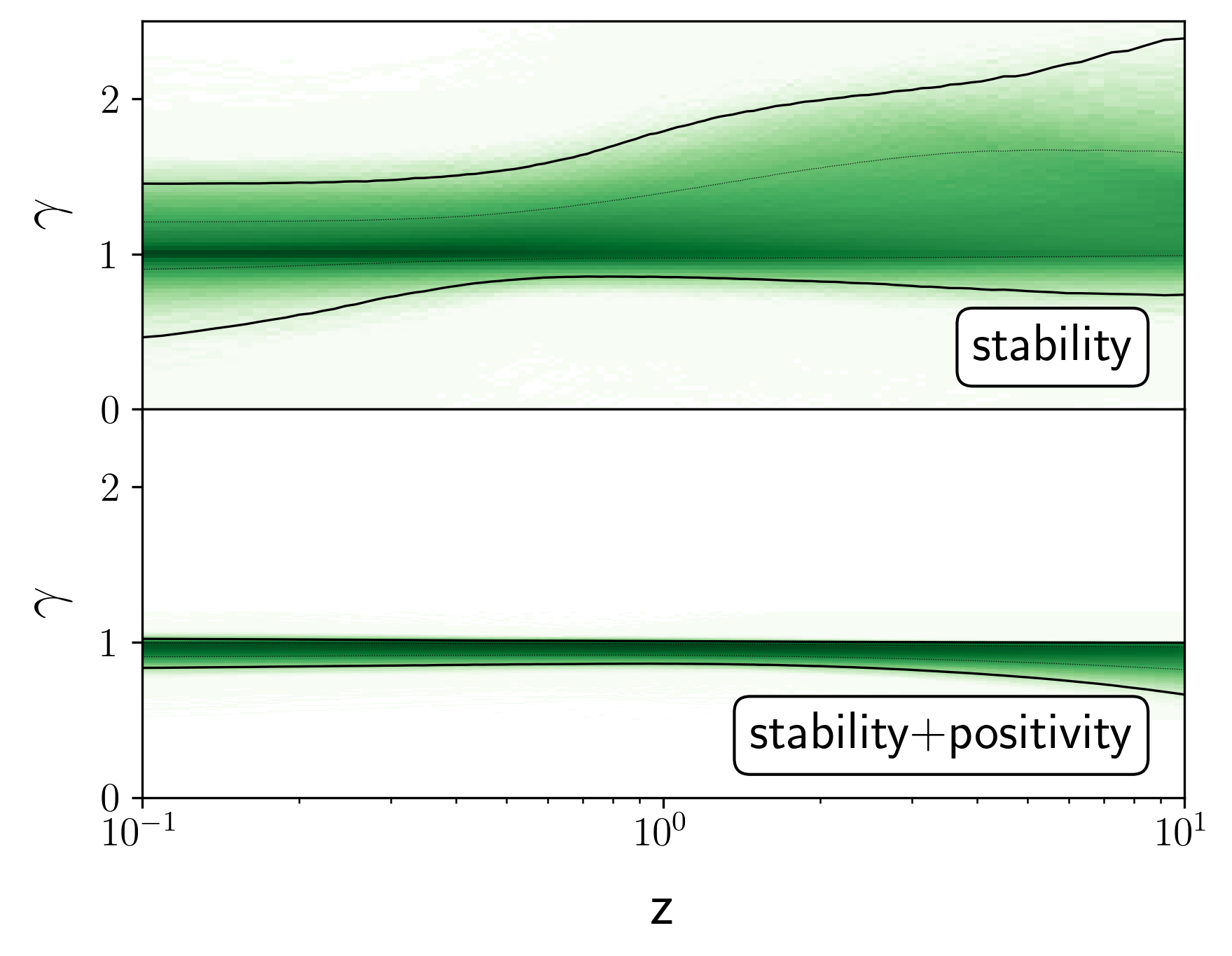}
    \includegraphics[width=0.48\textwidth]{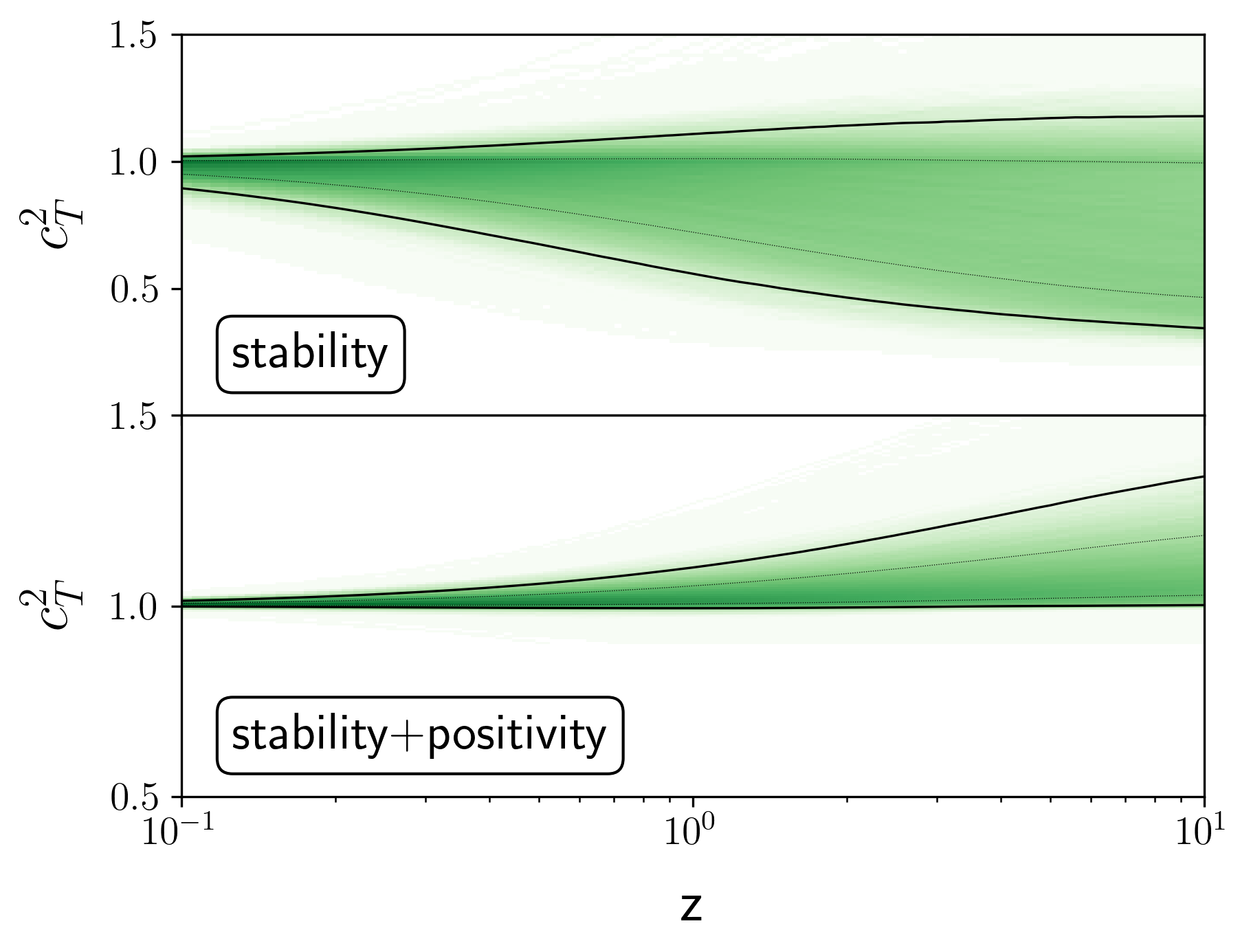}
     \caption{Theoretical distribution of the gravitational slip  $\gamma(z)$ ({\it left panel}) and the speed of tensors $c_T^2(z)$ ({\it right panel}) as functions of redshift. In the case of $\gamma$ we have chosen a representative scale $k=0.1\,{\rm Mpc}^{-1}$.  Color shading represents the density of function values at each redshift with dark color indicating high density. Solid (Dotted) lines plot the region where 95\% (68\%) of the function values reside.}
    \label{fig:slip_and_ct}
\end{figure}

We investigate this further by plotting  the sampled values of $\mu, \ \Sigma, \ \gamma$ in the continuous redshift range of $z\in[0.1,10]$, and the speed of tensor perturbation squared $c_T^2$, in Figure~\ref{fig:musgmgm} and~\ref{fig:slip_and_ct}.  Since  the results have negligible scale dependence, as can be noticed from the columns of Figure~\ref{fig:cloud}, we focus on the $k=0.1~\rm{Mpc}^{-1}$ case in Figure~\ref{fig:musgmgm} and Figure~\ref{fig:slip_and_ct}.
Looking at $\mu$ and $\Sigma$ separately, we see  that the main effect of positivity is that of ruling out most of the models with $\mu$ or $\Sigma$ significantly below unity while also tightening the constraint in general around $z\simeq1$. Comparing the plots of $\mu$ and $\Sigma$, we see that while the constrained $\mu(z)$ and $\Sigma(z)$ show some difference in the stability case, the difference is absent when positivity bounds are enforced, in agreement with the tight constraint on the gravitational slip, which we mentioned above and is now evident in Figure~\ref{fig:slip_and_ct} .
\begin{figure}
    \centering
\includegraphics[width=0.48\textwidth]{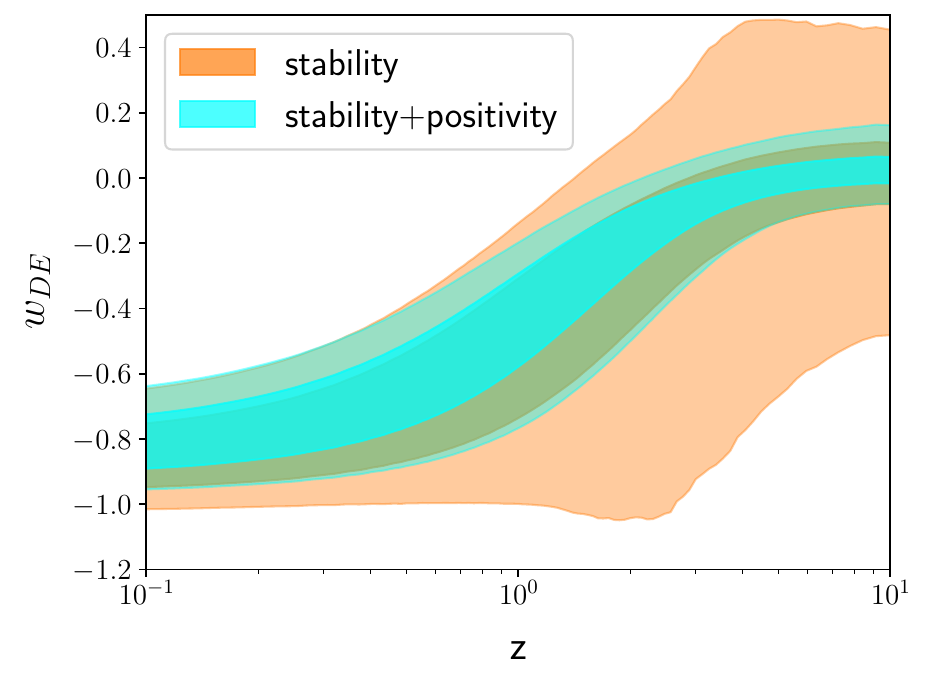}
    \includegraphics[width=0.465\textwidth]{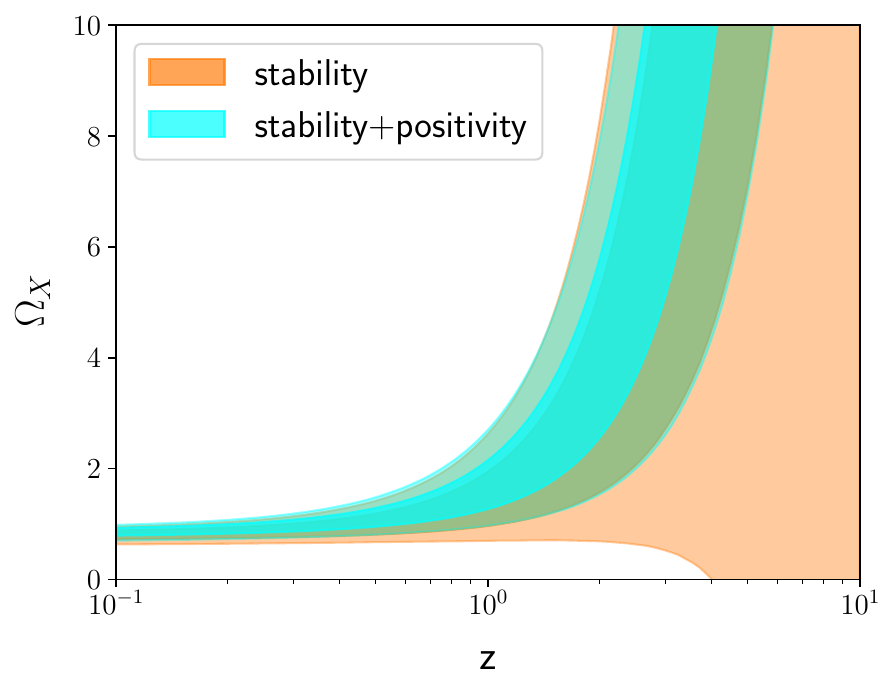}
    \caption{ Theoretical distribution of the effective dark energy equation of state $w_{\rm DE}(z)$ ({\it left panel}) and the fractional energy density $\Omega_X(z)$ ({\it right panel}). Shaded regions show the 68\% and 95\% confidence levels for the case with stability only (orange) and with stability+positivity(blue). }
    \label{fig:womg}
\end{figure}

Figure~\ref{fig:slip_and_ct} shows that positivity bounds strongly disfavor any model with a non-trivial gravitation slip, i.e. $\Psi\ne\Phi$; correspondingly, the distribution of $\gamma$ becomes a very narrow band around the $\gamma=1$ line, in stark contrast with the stability only results, which allows for considerable deviation from unity in the entire redshift range studied. From the theoretical perspective, within Horndeski gravity $\gamma\sim 1/c^2_T$ on large scales, while it can  receive contributions from the fifth-force on smaller scales~\cite{Pogosian:2016pwr}. Hence $\gamma\simeq 1$ has direct consequences on modifications to the speed of tensors and the non-minimal coupling. We can see this from the traceless part of spatial Einstein equation, written in the EFT convention of~\cite{Bellini:2014fua}, i.e. 
\begin{equation} \label{eq:einstein ij}
    \Psi-\Phi = \alpha_T \Phi + (\alpha_T-\alpha_M)H\delta\phi/\dot{\phi},
\end{equation}
where we have neglected matter anisotropic stress, $\delta\phi$ is the Horndeski field perturbation and $\alpha_T\equiv c^2_T-1, \alpha_M$ are respectively the tensor speed excess and running of effective Planck mass. For a general Horndeski model, there is no precise cancellation between the terms on the RHS of Eq.~\eqref{eq:einstein ij} thus $\Psi\simeq\Phi$ implies $\alpha_T\simeq\alpha_M\simeq0$, leading to nearly luminal $c_T^2$ \cite{Saltas:2014dha}. Our result confirms this theoretical observation in both panels of Figure~\ref{fig:slip_and_ct}, where positivity leads to stringent constraints on both the gravitational slip $\gamma$ and $c_T^2$. It is interesting to notice that in this case, the possible deviations of $\Sigma$ and $\mu$ from unity are attributable only to the kinetic braiding parameterized by the EFT function $\alpha_B$~\cite{Bellini:2014fua}. In other words, the models that live within these narrow bounds correspond to minimally-coupled Horndeski models with luminal tensors which, in the  language of action~\eqref{eq:Horndeskie}, means the $G_3(\phi,X)$ is the only non-trivial function.

The constraint on $\gamma$ loosens a bit and allows more models with $\gamma<1$ at high redshift, leading to the preference for $c_T^2\gtrsim1$ according to Eq.~\eqref{eq:einstein ij}. A preference for (super-)luminal $c_T^2$ was also found in~\cite{Melville:2019wyy}, where the authors investigate the impact of positivity bounds on the EFT parametrization $\{\alpha_M,\alpha_T,\alpha_K,\alpha_B\}$, as we review in Appendix \ref{App:horndeski}. It is interesting to note that Ref.~\cite{deRham:2021fpu} found a similar causality bound, for quartic theories, when adding in the formalism a positivity bound from a graviton-matter coupling of the form $\frac{1}{2M_{\mathrm{pl}}}h_{\mu \nu} T^{\mu \nu}$. However, such an interaction term is not included in our formalism, and therefore cannot be obviously linked to our findings. We leave the further investigation of the theoretical origin of the  improved constraints on $\gamma$ and $c_T^2$ for future work.   

Finally, in Figure~\ref{fig:womg} we show the effect of positivity bounds on the effective dark energy equation of state parameter $w_{\rm DE}$, defined by Eq.~\eqref{eq:wDE} and energy density $\Omega_X(z)$ defined in Eq.~\eqref{eq:omega_X}. The noisy 95\% edges of $w_{\rm DE}$ in the stability only case are due to $\rho_{\rm DE}$ crossing zero in some of the sampled models. Inclusion of positivity bounds significantly regularized the sampled models, removing the edge cases where $\rho_{\rm DE}$ crosses zero, hence the smoothed and tightened 95\% contour. On the other hand, at low-z, the shape of $w_{\rm DE}$ is dominated by the SNIa prior B4 (even though we inflated its error bar significantly), so the 68\% constraint is very similar between the stability and stability+positivity results. Positivity conditions do provide more constraining power at $z>1$ where data is absent and the posterior is driven by theoretical stability. As already discussed in~\cite{Raveri:2017qvt}, the sampled theories display a tracking behavior, with $w_{\rm DE}\simeq0$ in matter era; adding positivity only enhances this trend. This is also seen in the $\Omega_X$ plot, where adding positivity removes the models for which $\Omega_X$ decreases with redshift.

Before concluding this Section, let us note  that different representations can be used to sample models of the same EFT and a choice inevitably induces  a prior in the process. We based our main analysis on the EFT functions $\{\Omega,\Lambda,\gamma_1,\gamma_2,\gamma_3\}$; in this approach, the background is fully determined by the EFT functions, and can be solved for using~\eqref{eq:hubble}. In other words, a choice of the model consistently specifies background {\em and} perturbations. An alternative representation, used in some previous literature on positivity bounds~\cite{Melville:2019wyy,Kennedy:2020ehn,deRham:2021fpu}, is the one introduced in~\cite{Bellini:2014fua}, i.e. $\{\alpha_M,\alpha_T,\alpha_K,\alpha_B\}$; the latter needs to be used in combination with a separate choice for the background, i.e. one works in the so-called designer approach treating separately background and perturbations. While this may appear like a subtle difference, it can have some non-negligible impact on the outcome of the analysis. Ref.~\cite{Traykova:2021hbr} has raised concern about choosing the $\alpha$ representation over the covariant form. Importantly, as we discuss in  Appendix-\ref{App:horndeski},  direct parametrizations of the $\alpha$  functions risk to loose sight of any choice, e.g. shift-symmetry, imposed on the covariant action when the bounds are originally derived.

\section{Conclusions} \label{Sec:conclusion}
We have explored the effect of positivity bounds on the phenomenology of Horndeski gravity. Following the current literature,  we have assumed that the bounds derived on Minkowski background can be straightforwardly generalized  to the cosmological background. We have then applied a reconstruction method to translate the positivity bounds in terms of the covariant Horndeski functions $\{G_2,G_3,G_4.G_5\}$ to the formalism of EFTofDE which is at the basis of \texttt{EFTCAMB}, highlighting the assumptions made in this mapping. The outcome is a set of two conditions on the EFT functions $\{\Omega,\Lambda,\gamma_1,\gamma_2,\gamma_3\}$, to combine with the traditional stability conditions already present in \texttt{EFTCAMB}. As a result, we  could apply positivity bounds to Horndeski theory and study the corresponding cosmological phenomenology in a fully dynamical way, in which both the cosmological background and perturbations are consistently determined by the theory, i.e. specified by $\{\Omega,\Lambda,\gamma_1,\gamma_2,\gamma_3\}$. This sets our approach apart from previous studies of positivity bounds in Horndeski cosmology \cite{Melville:2019wyy,Kennedy:2020ehn,deRham:2021fpu,Traykova:2021hbr}, where either the theory sampling is limited to parameterized choices or the EFT functions depend on the background in a parameterized way rather than being dynamically determined by the theory to which positivity bounds are being imposed. In particular, we performed a comparison of our method and results with those of~\cite{Melville:2019wyy} in Appendix~\ref{App:horndeski}.

We have also introduced a new GP method for sampling the theory space, which gives more freedom with respect to that used in previous works~\cite{Raveri:2017qvt,Peirone:2017ywi,Espejo:2018hxa} while also having significantly fewer parameters and, therefore, leads to more representative samples. We find that the positivity conditions within Horndeski theory considerably tighten the theoretical bounds on $\Sigma,\mu$, with the most significant feature  being a further strengthening of the correlation $\mu\simeq\Sigma$, corresponding to  $\Phi/\Psi=\gamma\simeq 1$. We discuss how, in the context of Horndeski gravity, this is related to the corresponding tightening of the constraint on the speed of tensors into a narrow band around $c_T^2\simeq 1$. Within Horndeski theory, $\gamma\neq 1$ can be due to a modified speed of tensors and/or a fifth force, therefore the combination of stability and  positivity conditions, as formulated in this paper, induces a theoretical preference for cubic galileons. 

As we have discussed at length in the paper, there are still obstacles for a complete formulation of positivity bounds on cosmological backgrounds.  It will be interesting to repeat this analysis with a refined formulation of the bounds. Nevertheless, these results show the strong potential of such bounds in shaping
the viable parameter space of scalar-tensor theories. The constraints contributed by positivity conditions on the theoretical distributions of $w_{\rm DE}, \mu, \Sigma$ and $\gamma$ are impactful and particularly significant for the gravitational slip. While more work is needed in order to unveil the theoretical origin of the latter effect, it is interesting to notice that the $\gamma$ function will undergo scrutiny with the ongoing and upcoming Stage IV LSS surveys, such as Euclid~\cite{euclid}.

\section{Acknowledgements}
We thank Alice Garoffolo, Tanguy Grall for useful discussions. DdB, GY, AS acknowledge  support from the NWO and the Dutch Ministry of Education, Culture and Science (OCW) (through NWO VIDI Grant No. 2019/ENW/00678104 and ENW-XL Grant OCENW.XL21.XL21.025 DUSC) and from the D-ITP consortium.
NF is supported by the Italian Ministry of University and Research (MUR) through the Rita Levi Montalcini project ``Tests of gravity on cosmic scales" with reference PGR19ILFGP.
ISA is supported by Funda\c{c}\~{a}o para a Ci\^{e}ncia e a Tecnologia (FCT) through the PhD fellowship grant with ref. number 2020.07237.BD.
ISA and NF  also acknowledge the FCT project with ref. number PTDC/FIS-AST/0054/2021 and  the COST Action CosmoVerse, CA21136, supported by COST (European Cooperation in Science and Technology).
The Feynman diagrams have been produced with the online tool \url{https://feynman.aivazis.com/}.

\appendix

\section{Feynman diagrams and rules for the $\varphi\varphi\rightarrow\varphi\varphi$ scattering}\label{App:pos_bounds} 

\begin{figure}[h]
\includegraphics[scale=0.4]{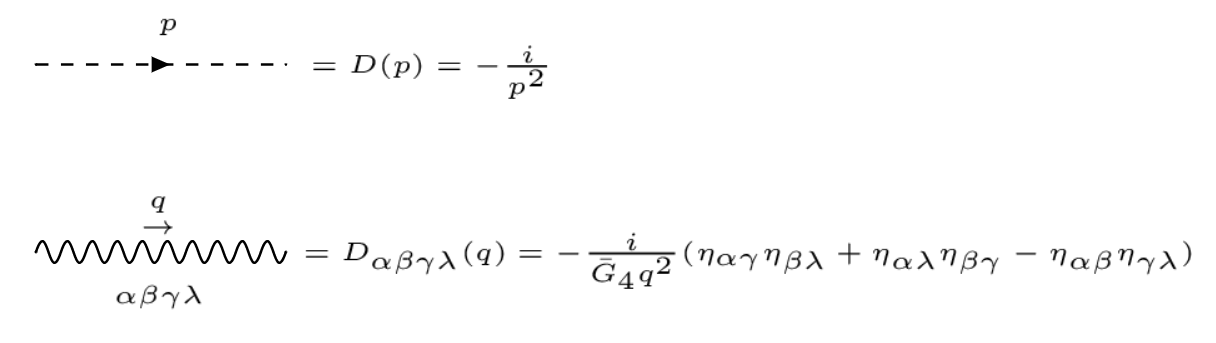}
\includegraphics[scale=0.4]{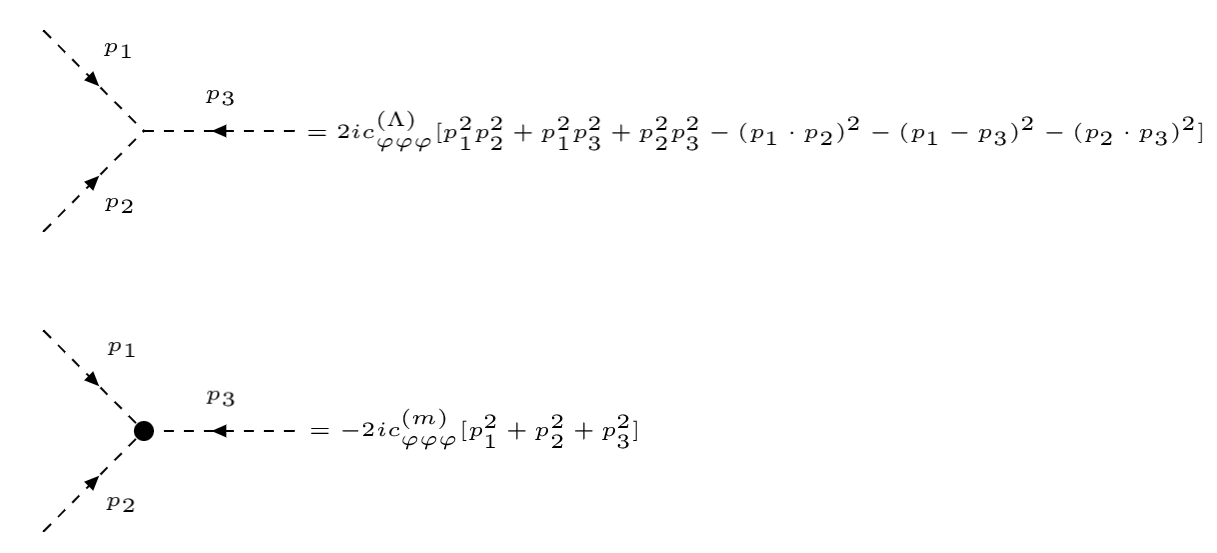}
\includegraphics[scale=0.4]{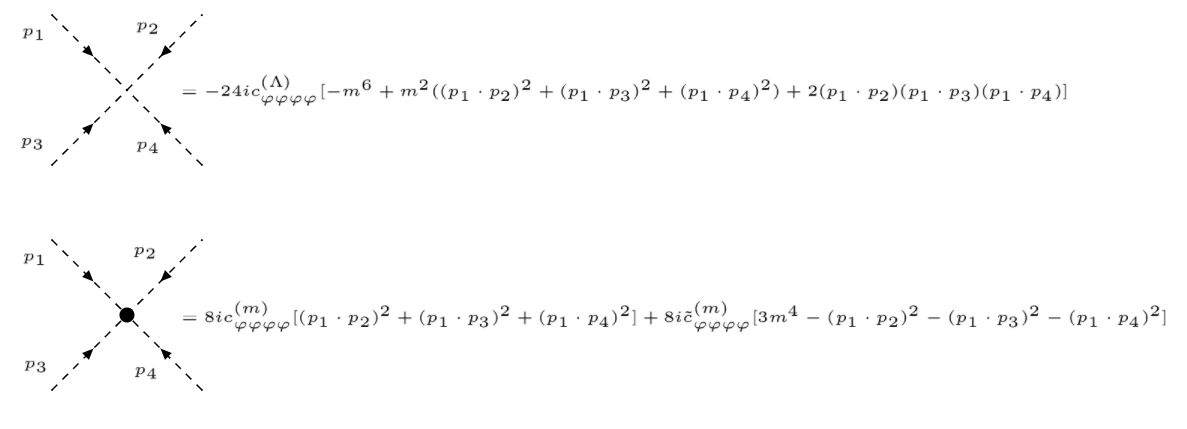}
\includegraphics[scale=0.4]{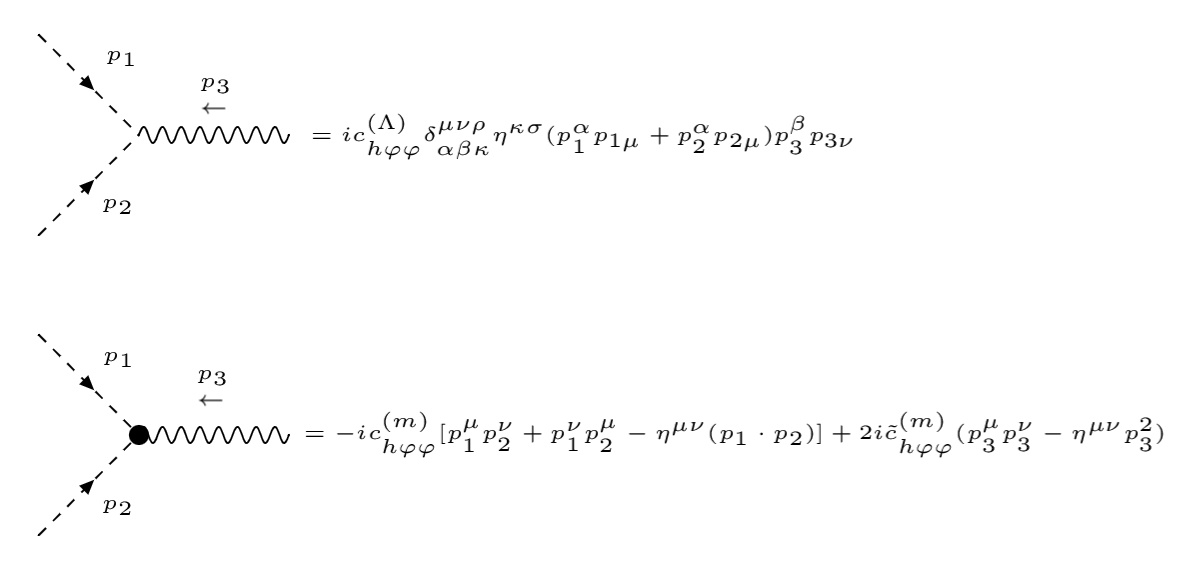}
\caption{Building blocks of the Feynman diagrams  for $\varphi \varphi \rightarrow \varphi \varphi$ tree-level scattering (to order $O(1/M_{\rm pl})$) in Horndeski gravity assuming a flat background. Dashed lines correspond to scalar fields, while wiggle lines indicate gravitons. Solid dots indicate sub-leading vertices ($\mathcal{O}(1/M_{\mathrm{pl}})$). The coefficients appearing in the Feynman rules are derived from the expressions of the vertices in~\eqref{eq:LagHorn}.} 
\label{fig:FeynHorn}
\end{figure}
\begin{figure}[h]
\includegraphics[scale=0.6]{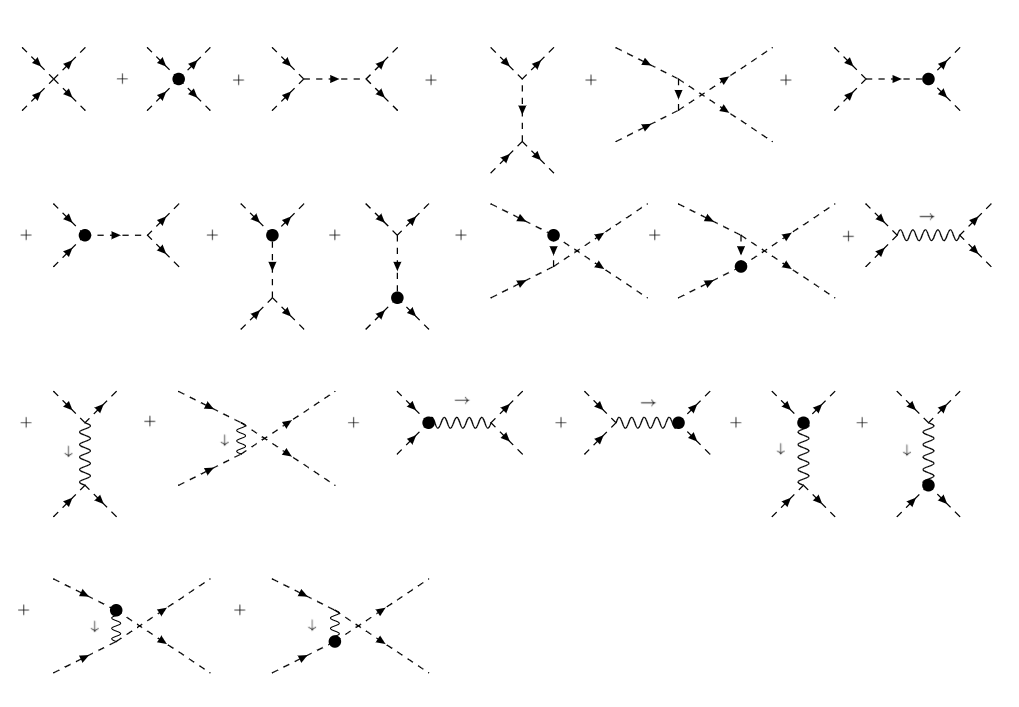}
\caption{Tree-level diagrams (some of the u-channel tree diagrams have crossing legs, which should not be confused with loops) for $\varphi \varphi \rightarrow \varphi \varphi$ scattering in Horndeski theory up to order $O(1/M_{\rm pl})$. The diagrams are organized in order of the type of vertex (including different channels) and in order of being leading or sub-leading order in $\mathcal{O}(1/M_{\mathrm{pl}})$. Dashed lines correspond to scalar fields, while wiggle lines indicate gravitons. Solid dots indicate sub-leading vertices ($\mathcal{O}(1/M_{\mathrm{pl}})$).
} 
\label{fig:Diagrams}
\end{figure}

In Section~\ref{Sec:poshorndeski} we have identified the relevant vertices for the $\varphi\varphi\rightarrow\varphi\varphi$ scattering process up to $O(1/M_{\rm pl})$. In order to calculate the corresponding scattering amplitudes, we use the following Feynman rules:
\begin{itemize}[leftmargin=*]
    \item external scalar fields correspond to $1$ in momentum space;
    \item external gravitons (on shell) contribute a polarization tensor $\epsilon^{\mu\nu}$ (and complex conjugate if the graviton is outgoing), which obeys $\epsilon^{\mu\nu}=\epsilon^{\nu\mu}$, $p_\mu \epsilon^{\mu\nu}(p)=0$, $\eta_{\mu\nu}\epsilon^{\mu\nu}=0$;
    \item internal scalar fields  with momentum $q$ correspond to the propagator $D(q)$;
    \item internal gravitons with momentum $q$ correspond to the propagator
    
    $D_{\alpha \beta \gamma \lambda}(q)=-\frac{i}{\bar{G}_4 q^2}\left(\eta_{\alpha\gamma}\eta_{\beta \lambda}+\eta_{\alpha\lambda}\eta_{\beta\gamma}-\eta_{\alpha\beta}\eta_{\gamma\lambda}\right)$.
\end{itemize}  
The rules are also depicted in Figure~\ref{fig:FeynHorn}. The total scattering amplitude is found by adding the amplitudes of the leading order diagrams included in Figure~\ref{fig:Diagrams}. 

Note that diagrams involving two subleading vertices have been ignored since these would correspond to $\mathcal{O}(1/M_{\mathrm{pl}}^2)$. 
For the calculation of the positivity bounds it is convenient to canonically normalize the scalar propagator with $\bar{G}_{2,\phi \phi} = \frac{m^2 M_{\mathrm{pl}}^2}{\Lambda_2^4} = m^2/H_0^2$ and $\bar{G}_{3,\phi} + \frac{1}{2}\bar{G}_{2,X} = 1/2$ such that $\mathcal{L}_{\varphi \varphi} = -\frac{1}{2} (\partial^\mu \varphi)(\partial_\mu \varphi)$ is the usual propagator. In the Feynman rules shown in Figure~\ref{fig:FeynHorn} this has been assumed, but not in the final positivity bounds. 
For Horndeski gravity without an explicit mass term $-\frac{1}{2}m^2 \varphi^2$ we have that $\bar{G}_{2,\phi \phi} = -m^2/H_0^2$. And we can leave out the assumption of canonical normalization $2\bar{G}_{3,\phi}+\bar{G}_{2,X} = 1$ by letting $2\bar{G}_{3,\phi} + \bar{G}_{2,X}>0$ be general \footnote{The field redefinition is then $\varphi \rightarrow \varphi \sqrt{2\bar{G}_{3,\phi}+\bar{G}_{2,X}}$ and for the mass $m^2 \rightarrow m^2/(\bar{G}_{2,X}+2\bar{G}_{3,\phi})$.}. Furthermore, we let $\bar{G}_4 \neq 0$ also be completely general.
\section{Reconstructed Horndeski gravity in EFTCAMB notation}\label{App:reconstruction}
Following the reconstruction procedure of
~\cite{Kennedy:2017sof}, given a set of EFT functions, we can determine the Horndeski models, i.e. the $G_i(
\phi, X)$ matching the dynamics at background and linear order. The result is:
\begin{align}
G_{2}(\phi,X) &= -M_\star^2 U(\phi) - \frac{1}{2}M_\star^2 Z(\phi)X + a_2(\phi) X^2 + \Delta G_2\,, \nonumber \\
G_3(\phi,X) &= b_1(\phi)X + \Delta G_3\,, \nonumber \\
G_4(\phi,X) &= \frac{1}{2}M_\star^2 F(\phi) + c_1(\phi)X + \Delta G_4 \,,\nonumber \\
G_5(\phi,X) &= \Delta G_5\,,
\end{align}
where  the functions of $\phi$ are combinations of the EFT functions as given in~\cite{Kennedy:2017sof}. Here we carefully adapt them to \texttt{EFTCAMB} notation, also in terms of the convention for derivatives, finding:
\begin{align}
U(\phi)&= -\frac{\Lambda}{m_0^2}+\frac{c}{m_0^2}-\frac{H_0^2}{2}\gamma_1-\frac{9H_0\mathcal{H}}{8a}\gamma_2-\frac{H_0\mathcal{H}}{8}\gamma_2^\prime+\frac{\dot{\mathcal{H}}+7\mathcal{H}^2}{2a^2}\gamma_3+\frac{\dot{\mathcal{H}}+7\mathcal{H}^2}{8a}\gamma_3^\prime+\frac{\mathcal{H}^2}{8}\gamma_3^{\prime\prime}\,, \nonumber \\
Z(\phi) &=\frac{2c}{m_0^6}-\frac{2H_0^2}{m_0^4}\gamma_1-\frac{3H_0\mathcal{H}}{2am_0^4}\gamma_2+\frac{H_0\mathcal{H}}{2m_0^4}\gamma^\prime_2 -\frac{2\left(\dot{\mathcal{H}}-\mathcal{H}^2\right)}{a^2m_0^4}\gamma_3-\frac{\left(\dot{\mathcal{H}}+\mathcal{H}^2\right)}{2am_0^4}\gamma_3^\prime-\frac{\mathcal{H}^2}{2m_0^4}\gamma_3^{\prime\prime} \,,\nonumber \\
a_2(\phi) &= \frac{H_0^2}{2m_0^6}\gamma_1-\frac{3H_0\mathcal{H}}{8am_0^6}\gamma_2+\frac{H_0\mathcal{H}}{8m_0^6}\gamma_2^\prime+\frac{\dot{\mathcal{H}}-\frac{5}{2}\mathcal{H}^2}{2a^2m_0^6}\gamma_3-\frac{\dot{\mathcal{H}}-\mathcal{H}^2}{8am_0^6}\gamma_3^\prime- \frac{\mathcal{H}^2}{8m_0^6}\gamma_3^{\prime\prime}\,, \nonumber \\
b_1(\phi)&=\frac{\mathcal{H}}{am_0^4}\gamma_3-\frac{\mathcal{H}}{2m_0^4}\gamma_3^\prime+\frac{H_0}{2m_0^4}\gamma_2 \,, \nonumber \\
F(\phi) &= 1+\Omega+\frac{1}{2}\gamma_3\,, \nonumber \\
c_1(\phi) &= \frac{\gamma_3}{4m_0^2}\,,
\end{align}
where a prime denotes derivation w.r.t. the scale factor,  a dot derivation w.r.t. conformal time and $m_0$ is the Planck mass.

\section{Positivity constraints on the $\{\alpha_M,\alpha_T,\alpha_K,\alpha_B\}$ parameterized Horndeski theory}\label{App:horndeski}

\begin{figure}[ht!]
    \centering
    \includegraphics[width=0.8\linewidth]{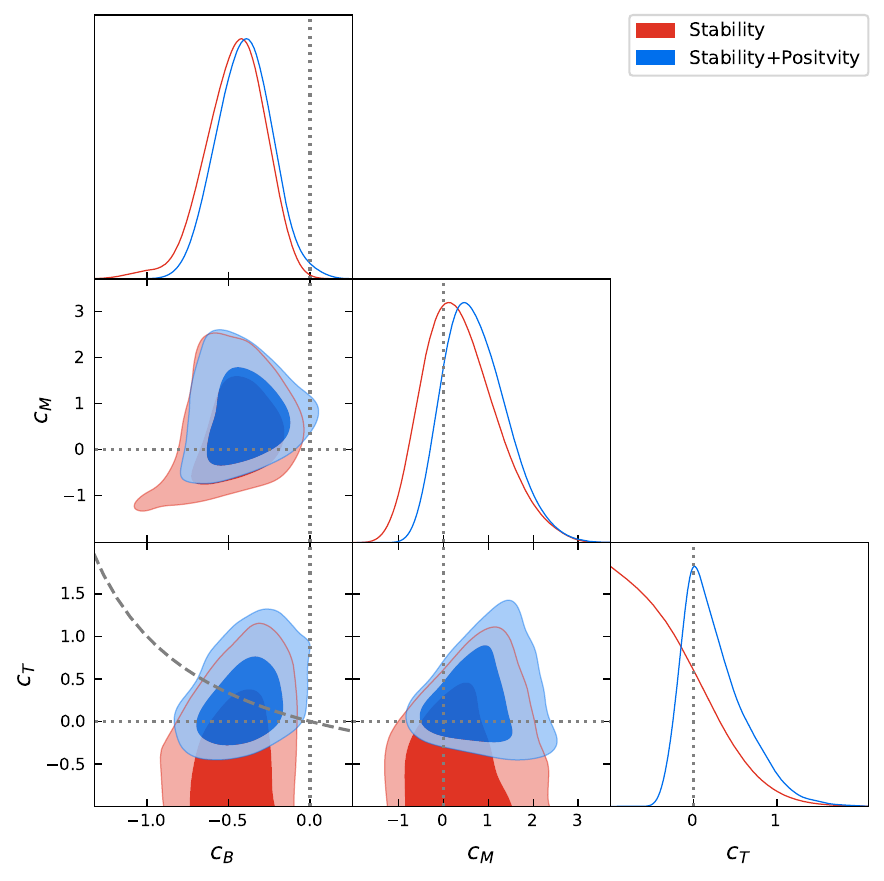}
    \caption{68\% and 95\% posterior distributions of the Horndeski parameters $\{c_T,c_B,c_M\}$. Gray dotted lines indicate the GR prediction $\alpha_i=0$. Gray dashed line marks the boundary of $c_B \leq 2c_T /(1 + c_T )$, i.e. the positivity bound reported by MN19. }
    \label{fig:alphaDE_trig}
\end{figure}
Ref.~\cite{Melville:2019wyy} (hereafter MN19) has studied positivity bounds in the Horndeski theory using the $\{\alpha_M,\alpha_T,\alpha_K,\alpha_B\}$ parameterization assuming a standard $\Lambda$CDM background and
\begin{equation}\label{eq:alphaDE}
    \alpha_i(a) = c_i\Omega_{\rm{DE}}(a),\quad i\in\{M,T,B,K\}
\end{equation}
where $c_i$'s are constants and $\Omega_{\rm{DE}}(a)$ is the energy fraction of dark energy at scale factor $a$. For comparison, we implement the parameterization~\eqref{eq:alphaDE} in \texttt{EFTCAMB} and present in this Appendix the stability and positivity constraints on the $c_i$'s. Following Ref.\cite{Melville:2019wyy} we set $c_K=0$ and use Planck 2018 CMB high-$l$ lite and low-$l$ temperature and polarization data as well as CMB lensing \cite{Planck:2018vyg, Planck:2018lbu}, baryon acoustic oscillation (BAO) measurements from 6dF \cite{Beutler:2011hx}, MGS \cite{Ross:2014qpa} and SDSS DR12 final consensus, including redshift space distortion (RSD) constraints on $f\sigma_8$ \cite{BOSS:2016wmc}. We also include the Pantheon SNIa data \cite{Pan-STARRS1:2017jku} to further constrain the background evolution. In the MCMC analysis we vary the 6 standard $\Lambda$CDM parameters $\{\Omega_c h^2, \Omega_b h^2, H_0, \ln(10^9A_s), n_s, \tau_{re}\}$ plus $\{c_M, c_T, c_B\}$. For each model we check for (ghost and gradient) \textit{Stability} and/or \textit{Positivity} (Eqs.~\eqref{eq:posbound1}, \eqref{eq:posbound2}). Before presenting the results, it is important to stress that our approach has some differences compared with MN19: we include SNIa observations and use more recent (and thus more constraining) CMB, BAO and RSD data, removing any data that directly constrain the shape of the matter power spectrum because reliability of non-linear correction is not clear in general modified gravity/dark energy models;  due to different conventions, the $\alpha_B$ in \texttt{EFTCAMB} is equal to $-\alpha_B$ in MN19; and, finally the most relevant difference is in the positivity bounds conditions themselves. Namely, MN19 used the bounds derived from a specific shift-symmetric Lagrangian (not necessarily corresponding to~\eqref{eq:alphaDE}) while we applied general Horndeski positivity bounds to the reconstructed covariant theory (guaranteed to reproduce the given background and~\eqref{eq:alphaDE}) described in Section~\ref{sec:code_notation}. While it may appear subtle, this is a fundamental difference. 
Figure \ref{fig:alphaDE_trig} reports the posterior results for $\{c_M, c_T, c_B\}$. The \textit{Stability} result is similar to MN19, particularly the degeneracy direction in the $c_M-c_B$ plane coincides. We obtain a tighter constraint on $c_B$, which is probably due to the improved dataset and numeric difference in code for edge cases. Because of the difference in positivity bounds discussed above, we find milder constraints from positivity. Despite the difference, we observe preference for (super-)luminal gravitational waves and $c_M>0$ with positivity, similarly to MN19 just somewhat milder. Moreover, it is very interesting that positivity prefers (super-)luminal tensor speed for both the parameterized case of~\eqref{eq:alphaDE} and the parameterization-independent case studied in the main text. 

\bibliography{References}

\begin{thebibliography}{88}%
\makeatletter
\providecommand \@ifxundefined [1]{%
 \@ifx{#1\undefined}
}%
\providecommand \@ifnum [1]{%
 \ifnum #1\expandafter \@firstoftwo
 \else \expandafter \@secondoftwo
 \fi
}%
\providecommand \@ifx [1]{%
 \ifx #1\expandafter \@firstoftwo
 \else \expandafter \@secondoftwo
 \fi
}%
\providecommand \natexlab [1]{#1}%
\providecommand \enquote  [1]{``#1''}%
\providecommand \bibnamefont  [1]{#1}%
\providecommand \bibfnamefont [1]{#1}%
\providecommand \citenamefont [1]{#1}%
\providecommand \href@noop [0]{\@secondoftwo}%
\providecommand \href [0]{\begingroup \@sanitize@url \@href}%
\providecommand \@href[1]{\@@startlink{#1}\@@href}%
\providecommand \@@href[1]{\endgroup#1\@@endlink}%
\providecommand \@sanitize@url [0]{\catcode `\\12\catcode `\$12\catcode
  `\&12\catcode `\#12\catcode `\^12\catcode `\_12\catcode `\%12\relax}%
\providecommand \@@startlink[1]{}%
\providecommand \@@endlink[0]{}%
\providecommand \url  [0]{\begingroup\@sanitize@url \@url }%
\providecommand \@url [1]{\endgroup\@href {#1}{\urlprefix }}%
\providecommand \urlprefix  [0]{URL }%
\providecommand \Eprint [0]{\href }%
\providecommand \doibase [0]{http://dx.doi.org/}%
\providecommand \selectlanguage [0]{\@gobble}%
\providecommand \bibinfo  [0]{\@secondoftwo}%
\providecommand \bibfield  [0]{\@secondoftwo}%
\providecommand \translation [1]{[#1]}%
\providecommand \BibitemOpen [0]{}%
\providecommand \bibitemStop [0]{}%
\providecommand \bibitemNoStop [0]{.\EOS\space}%
\providecommand \EOS [0]{\spacefactor3000\relax}%
\providecommand \BibitemShut  [1]{\csname bibitem#1\endcsname}%
\let\auto@bib@innerbib\@empty
\bibitem [{\citenamefont {Weinberg}(1980)}]{Weinberg:1980wa}%
  \BibitemOpen
  \bibfield  {author} {\bibinfo {author} {\bibfnamefont {S.}~\bibnamefont
  {Weinberg}},\ }\href {\doibase 10.1016/0370-2693(80)90660-7} {\bibfield
  {journal} {\bibinfo  {journal} {Phys. Lett. B}\ }\textbf {\bibinfo {volume}
  {91}},\ \bibinfo {pages} {51} (\bibinfo {year} {1980})}\BibitemShut {NoStop}%
\bibitem [{\citenamefont {Wilson}(1983)}]{Wilson:1983xri}%
  \BibitemOpen
  \bibfield  {author} {\bibinfo {author} {\bibfnamefont {K.~G.}\ \bibnamefont
  {Wilson}},\ }\href {\doibase 10.1103/RevModPhys.55.583} {\bibfield  {journal}
  {\bibinfo  {journal} {Rev. Mod. Phys.}\ }\textbf {\bibinfo {volume} {55}},\
  \bibinfo {pages} {583} (\bibinfo {year} {1983})}\BibitemShut {NoStop}%
\bibitem [{\citenamefont {Penco}(2020)}]{Penco:2020kvy}%
  \BibitemOpen
  \bibfield  {author} {\bibinfo {author} {\bibfnamefont {R.}~\bibnamefont
  {Penco}},\ }\href@noop {} {\  (\bibinfo {year} {2020})},\ \Eprint
  {http://arxiv.org/abs/2006.16285} {arXiv:2006.16285 [hep-th]} \BibitemShut
  {NoStop}%
\bibitem [{\citenamefont {Weinberg}(2021)}]{Weinberg:2021exr}%
  \BibitemOpen
  \bibfield  {author} {\bibinfo {author} {\bibfnamefont {S.}~\bibnamefont
  {Weinberg}},\ }\href {\doibase 10.1140/epjh/s13129-021-00004-x} {\bibfield
  {journal} {\bibinfo  {journal} {Eur. Phys. J. H}\ }\textbf {\bibinfo {volume}
  {46}},\ \bibinfo {pages} {6} (\bibinfo {year} {2021})},\ \Eprint
  {http://arxiv.org/abs/2101.04241} {arXiv:2101.04241 [hep-th]} \BibitemShut
  {NoStop}%
\bibitem [{\citenamefont {Baumgart}\ \emph {et~al.}(2022)\citenamefont
  {Baumgart} \emph {et~al.}}]{Baumgart:2022yty}%
  \BibitemOpen
  \bibfield  {author} {\bibinfo {author} {\bibfnamefont {M.}~\bibnamefont
  {Baumgart}} \emph {et~al.},\ }in\ \href@noop {} {\emph {\bibinfo {booktitle}
  {{Snowmass 2021}}}}\ (\bibinfo {year} {2022})\ \Eprint
  {http://arxiv.org/abs/2210.03199} {arXiv:2210.03199 [hep-ph]} \BibitemShut
  {NoStop}%
\bibitem [{\citenamefont {Weinberg}(2008)}]{Weinberg:2008hq}%
  \BibitemOpen
  \bibfield  {author} {\bibinfo {author} {\bibfnamefont {S.}~\bibnamefont
  {Weinberg}},\ }\href {\doibase 10.1103/PhysRevD.77.123541} {\bibfield
  {journal} {\bibinfo  {journal} {Phys. Rev. D}\ }\textbf {\bibinfo {volume}
  {77}},\ \bibinfo {pages} {123541} (\bibinfo {year} {2008})},\ \Eprint
  {http://arxiv.org/abs/0804.4291} {arXiv:0804.4291 [hep-th]} \BibitemShut
  {NoStop}%
\bibitem [{\citenamefont {Cheung}\ \emph {et~al.}(2008)\citenamefont {Cheung},
  \citenamefont {Creminelli}, \citenamefont {Fitzpatrick}, \citenamefont
  {Kaplan},\ and\ \citenamefont {Senatore}}]{Cheung:2007st}%
  \BibitemOpen
  \bibfield  {author} {\bibinfo {author} {\bibfnamefont {C.}~\bibnamefont
  {Cheung}}, \bibinfo {author} {\bibfnamefont {P.}~\bibnamefont {Creminelli}},
  \bibinfo {author} {\bibfnamefont {A.~L.}\ \bibnamefont {Fitzpatrick}},
  \bibinfo {author} {\bibfnamefont {J.}~\bibnamefont {Kaplan}}, \ and\ \bibinfo
  {author} {\bibfnamefont {L.}~\bibnamefont {Senatore}},\ }\href {\doibase
  10.1088/1126-6708/2008/03/014} {\bibfield  {journal} {\bibinfo  {journal}
  {JHEP}\ }\textbf {\bibinfo {volume} {03}},\ \bibinfo {pages} {014} (\bibinfo
  {year} {2008})},\ \Eprint {http://arxiv.org/abs/0709.0293} {arXiv:0709.0293
  [hep-th]} \BibitemShut {NoStop}%
\bibitem [{\citenamefont {Senatore}\ and\ \citenamefont
  {Zaldarriaga}(2012)}]{Senatore:2010wk}%
  \BibitemOpen
  \bibfield  {author} {\bibinfo {author} {\bibfnamefont {L.}~\bibnamefont
  {Senatore}}\ and\ \bibinfo {author} {\bibfnamefont {M.}~\bibnamefont
  {Zaldarriaga}},\ }\href {\doibase 10.1007/JHEP04(2012)024} {\bibfield
  {journal} {\bibinfo  {journal} {JHEP}\ }\textbf {\bibinfo {volume} {04}},\
  \bibinfo {pages} {024} (\bibinfo {year} {2012})},\ \Eprint
  {http://arxiv.org/abs/1009.2093} {arXiv:1009.2093 [hep-th]} \BibitemShut
  {NoStop}%
\bibitem [{\citenamefont {Baumann}\ \emph {et~al.}(2012)\citenamefont
  {Baumann}, \citenamefont {Nicolis}, \citenamefont {Senatore},\ and\
  \citenamefont {Zaldarriaga}}]{Baumann:2010tm}%
  \BibitemOpen
  \bibfield  {author} {\bibinfo {author} {\bibfnamefont {D.}~\bibnamefont
  {Baumann}}, \bibinfo {author} {\bibfnamefont {A.}~\bibnamefont {Nicolis}},
  \bibinfo {author} {\bibfnamefont {L.}~\bibnamefont {Senatore}}, \ and\
  \bibinfo {author} {\bibfnamefont {M.}~\bibnamefont {Zaldarriaga}},\ }\href
  {\doibase 10.1088/1475-7516/2012/07/051} {\bibfield  {journal} {\bibinfo
  {journal} {JCAP}\ }\textbf {\bibinfo {volume} {07}},\ \bibinfo {pages} {051}
  (\bibinfo {year} {2012})},\ \Eprint {http://arxiv.org/abs/1004.2488}
  {arXiv:1004.2488 [astro-ph.CO]} \BibitemShut {NoStop}%
\bibitem [{\citenamefont {Carrasco}\ \emph {et~al.}(2012)\citenamefont
  {Carrasco}, \citenamefont {Hertzberg},\ and\ \citenamefont
  {Senatore}}]{Carrasco:2012cv}%
  \BibitemOpen
  \bibfield  {author} {\bibinfo {author} {\bibfnamefont {J.~J.~M.}\
  \bibnamefont {Carrasco}}, \bibinfo {author} {\bibfnamefont {M.~P.}\
  \bibnamefont {Hertzberg}}, \ and\ \bibinfo {author} {\bibfnamefont
  {L.}~\bibnamefont {Senatore}},\ }\href {\doibase 10.1007/JHEP09(2012)082}
  {\bibfield  {journal} {\bibinfo  {journal} {JHEP}\ }\textbf {\bibinfo
  {volume} {09}},\ \bibinfo {pages} {082} (\bibinfo {year} {2012})},\ \Eprint
  {http://arxiv.org/abs/1206.2926} {arXiv:1206.2926 [astro-ph.CO]} \BibitemShut
  {NoStop}%
\bibitem [{\citenamefont {Hertzberg}(2014)}]{Hertzberg:2012qn}%
  \BibitemOpen
  \bibfield  {author} {\bibinfo {author} {\bibfnamefont {M.~P.}\ \bibnamefont
  {Hertzberg}},\ }\href {\doibase 10.1103/PhysRevD.89.043521} {\bibfield
  {journal} {\bibinfo  {journal} {Phys. Rev. D}\ }\textbf {\bibinfo {volume}
  {89}},\ \bibinfo {pages} {043521} (\bibinfo {year} {2014})},\ \Eprint
  {http://arxiv.org/abs/1208.0839} {arXiv:1208.0839 [astro-ph.CO]} \BibitemShut
  {NoStop}%
\bibitem [{\citenamefont {Bloomfield}\ \emph {et~al.}(2013)\citenamefont
  {Bloomfield}, \citenamefont {Flanagan}, \citenamefont {Park},\ and\
  \citenamefont {Watson}}]{Bloomfield:2012ff}%
  \BibitemOpen
  \bibfield  {author} {\bibinfo {author} {\bibfnamefont {J.~K.}\ \bibnamefont
  {Bloomfield}}, \bibinfo {author} {\bibfnamefont {E.~E.}\ \bibnamefont
  {Flanagan}}, \bibinfo {author} {\bibfnamefont {M.}~\bibnamefont {Park}}, \
  and\ \bibinfo {author} {\bibfnamefont {S.}~\bibnamefont {Watson}},\ }\href
  {\doibase 10.1088/1475-7516/2013/08/010} {\bibfield  {journal} {\bibinfo
  {journal} {JCAP}\ }\textbf {\bibinfo {volume} {1308}},\ \bibinfo {pages}
  {010} (\bibinfo {year} {2013})},\ \Eprint {http://arxiv.org/abs/1211.7054}
  {arXiv:1211.7054 [astro-ph.CO]} \BibitemShut {NoStop}%
\bibitem [{\citenamefont {Gubitosi}\ \emph {et~al.}(2013)\citenamefont
  {Gubitosi}, \citenamefont {Piazza},\ and\ \citenamefont
  {Vernizzi}}]{Gubitosi:2012hu}%
  \BibitemOpen
  \bibfield  {author} {\bibinfo {author} {\bibfnamefont {G.}~\bibnamefont
  {Gubitosi}}, \bibinfo {author} {\bibfnamefont {F.}~\bibnamefont {Piazza}}, \
  and\ \bibinfo {author} {\bibfnamefont {F.}~\bibnamefont {Vernizzi}},\ }\href
  {\doibase 10.1088/1475-7516/2013/02/032} {\bibfield  {journal} {\bibinfo
  {journal} {JCAP}\ }\textbf {\bibinfo {volume} {1302}},\ \bibinfo {pages}
  {032} (\bibinfo {year} {2013})},\ \bibinfo {note} {[JCAP1302,032(2013)]},\
  \Eprint {http://arxiv.org/abs/1210.0201} {arXiv:1210.0201 [hep-th]}
  \BibitemShut {NoStop}%
\bibitem [{\citenamefont {Piazza}\ and\ \citenamefont
  {Vernizzi}(2013)}]{Piazza:2013coa}%
  \BibitemOpen
  \bibfield  {author} {\bibinfo {author} {\bibfnamefont {F.}~\bibnamefont
  {Piazza}}\ and\ \bibinfo {author} {\bibfnamefont {F.}~\bibnamefont
  {Vernizzi}},\ }\href {\doibase 10.1088/0264-9381/30/21/214007} {\bibfield
  {journal} {\bibinfo  {journal} {Class. Quant. Grav.}\ }\textbf {\bibinfo
  {volume} {30}},\ \bibinfo {pages} {214007} (\bibinfo {year} {2013})},\
  \Eprint {http://arxiv.org/abs/1307.4350} {arXiv:1307.4350 [hep-th]}
  \BibitemShut {NoStop}%
\bibitem [{\citenamefont {Frusciante}\ and\ \citenamefont
  {Perenon}(2020)}]{Frusciante:2019xia}%
  \BibitemOpen
  \bibfield  {author} {\bibinfo {author} {\bibfnamefont {N.}~\bibnamefont
  {Frusciante}}\ and\ \bibinfo {author} {\bibfnamefont {L.}~\bibnamefont
  {Perenon}},\ }\href {\doibase 10.1016/j.physrep.2020.02.004} {\bibfield
  {journal} {\bibinfo  {journal} {Phys. Rept.}\ }\textbf {\bibinfo {volume}
  {857}},\ \bibinfo {pages} {1} (\bibinfo {year} {2020})},\ \Eprint
  {http://arxiv.org/abs/1907.03150} {arXiv:1907.03150 [astro-ph.CO]}
  \BibitemShut {NoStop}%
\bibitem [{\citenamefont {Adams}\ \emph {et~al.}(2006)\citenamefont {Adams},
  \citenamefont {Arkani-Hamed}, \citenamefont {Dubovsky}, \citenamefont
  {Nicolis},\ and\ \citenamefont {Rattazzi}}]{Adams:2006sv}%
  \BibitemOpen
  \bibfield  {author} {\bibinfo {author} {\bibfnamefont {A.}~\bibnamefont
  {Adams}}, \bibinfo {author} {\bibfnamefont {N.}~\bibnamefont {Arkani-Hamed}},
  \bibinfo {author} {\bibfnamefont {S.}~\bibnamefont {Dubovsky}}, \bibinfo
  {author} {\bibfnamefont {A.}~\bibnamefont {Nicolis}}, \ and\ \bibinfo
  {author} {\bibfnamefont {R.}~\bibnamefont {Rattazzi}},\ }\href {\doibase
  10.1088/1126-6708/2006/10/014} {\bibfield  {journal} {\bibinfo  {journal}
  {JHEP}\ }\textbf {\bibinfo {volume} {10}},\ \bibinfo {pages} {014} (\bibinfo
  {year} {2006})},\ \Eprint {http://arxiv.org/abs/hep-th/0602178}
  {arXiv:hep-th/0602178} \BibitemShut {NoStop}%
\bibitem [{\citenamefont {Joyce}\ \emph {et~al.}(2015)\citenamefont {Joyce},
  \citenamefont {Jain}, \citenamefont {Khoury},\ and\ \citenamefont
  {Trodden}}]{Joyce:2014kja}%
  \BibitemOpen
  \bibfield  {author} {\bibinfo {author} {\bibfnamefont {A.}~\bibnamefont
  {Joyce}}, \bibinfo {author} {\bibfnamefont {B.}~\bibnamefont {Jain}},
  \bibinfo {author} {\bibfnamefont {J.}~\bibnamefont {Khoury}}, \ and\ \bibinfo
  {author} {\bibfnamefont {M.}~\bibnamefont {Trodden}},\ }\href {\doibase
  10.1016/j.physrep.2014.12.002} {\bibfield  {journal} {\bibinfo  {journal}
  {Phys. Rept.}\ }\textbf {\bibinfo {volume} {568}},\ \bibinfo {pages} {1}
  (\bibinfo {year} {2015})},\ \Eprint {http://arxiv.org/abs/1407.0059}
  {arXiv:1407.0059 [astro-ph.CO]} \BibitemShut {NoStop}%
\bibitem [{\citenamefont {Nicolis}\ \emph {et~al.}(2010)\citenamefont
  {Nicolis}, \citenamefont {Rattazzi},\ and\ \citenamefont
  {Trincherini}}]{Nicolis:2009qm}%
  \BibitemOpen
  \bibfield  {author} {\bibinfo {author} {\bibfnamefont {A.}~\bibnamefont
  {Nicolis}}, \bibinfo {author} {\bibfnamefont {R.}~\bibnamefont {Rattazzi}}, \
  and\ \bibinfo {author} {\bibfnamefont {E.}~\bibnamefont {Trincherini}},\
  }\href {\doibase 10.1007/JHEP05(2010)095, 10.1007/JHEP11(2011)128} {\bibfield
   {journal} {\bibinfo  {journal} {JHEP}\ }\textbf {\bibinfo {volume} {05}},\
  \bibinfo {pages} {095} (\bibinfo {year} {2010})},\ \bibinfo {note} {[Erratum:
  JHEP11,128(2011)]},\ \Eprint {http://arxiv.org/abs/0912.4258}
  {arXiv:0912.4258 [hep-th]} \BibitemShut {NoStop}%
\bibitem [{\citenamefont {Baumann}\ \emph {et~al.}(2016)\citenamefont
  {Baumann}, \citenamefont {Green}, \citenamefont {Lee},\ and\ \citenamefont
  {Porto}}]{Baumann:2015nta}%
  \BibitemOpen
  \bibfield  {author} {\bibinfo {author} {\bibfnamefont {D.}~\bibnamefont
  {Baumann}}, \bibinfo {author} {\bibfnamefont {D.}~\bibnamefont {Green}},
  \bibinfo {author} {\bibfnamefont {H.}~\bibnamefont {Lee}}, \ and\ \bibinfo
  {author} {\bibfnamefont {R.~A.}\ \bibnamefont {Porto}},\ }\href {\doibase
  10.1103/PhysRevD.93.023523} {\bibfield  {journal} {\bibinfo  {journal} {Phys.
  Rev. D}\ }\textbf {\bibinfo {volume} {93}},\ \bibinfo {pages} {023523}
  (\bibinfo {year} {2016})},\ \Eprint {http://arxiv.org/abs/1502.07304}
  {arXiv:1502.07304 [hep-th]} \BibitemShut {NoStop}%
\bibitem [{\citenamefont {Bellazzini}\ \emph {et~al.}(2016)\citenamefont
  {Bellazzini}, \citenamefont {Cheung},\ and\ \citenamefont
  {Remmen}}]{Bellazzini:2015cra}%
  \BibitemOpen
  \bibfield  {author} {\bibinfo {author} {\bibfnamefont {B.}~\bibnamefont
  {Bellazzini}}, \bibinfo {author} {\bibfnamefont {C.}~\bibnamefont {Cheung}},
  \ and\ \bibinfo {author} {\bibfnamefont {G.~N.}\ \bibnamefont {Remmen}},\
  }\href {\doibase 10.1103/PhysRevD.93.064076} {\bibfield  {journal} {\bibinfo
  {journal} {Phys. Rev. D}\ }\textbf {\bibinfo {volume} {93}},\ \bibinfo
  {pages} {064076} (\bibinfo {year} {2016})},\ \Eprint
  {http://arxiv.org/abs/1509.00851} {arXiv:1509.00851 [hep-th]} \BibitemShut
  {NoStop}%
\bibitem [{\citenamefont {Cheung}\ and\ \citenamefont
  {Remmen}(2016)}]{Cheung:2016yqr}%
  \BibitemOpen
  \bibfield  {author} {\bibinfo {author} {\bibfnamefont {C.}~\bibnamefont
  {Cheung}}\ and\ \bibinfo {author} {\bibfnamefont {G.~N.}\ \bibnamefont
  {Remmen}},\ }\href {\doibase 10.1007/JHEP04(2016)002} {\bibfield  {journal}
  {\bibinfo  {journal} {JHEP}\ }\textbf {\bibinfo {volume} {04}},\ \bibinfo
  {pages} {002} (\bibinfo {year} {2016})},\ \Eprint
  {http://arxiv.org/abs/1601.04068} {arXiv:1601.04068 [hep-th]} \BibitemShut
  {NoStop}%
\bibitem [{\citenamefont {Cheung}\ and\ \citenamefont
  {Remmen}(2017)}]{Cheung:2016wjt}%
  \BibitemOpen
  \bibfield  {author} {\bibinfo {author} {\bibfnamefont {C.}~\bibnamefont
  {Cheung}}\ and\ \bibinfo {author} {\bibfnamefont {G.~N.}\ \bibnamefont
  {Remmen}},\ }\href {\doibase 10.1103/PhysRevLett.118.051601} {\bibfield
  {journal} {\bibinfo  {journal} {Phys. Rev. Lett.}\ }\textbf {\bibinfo
  {volume} {118}},\ \bibinfo {pages} {051601} (\bibinfo {year} {2017})},\
  \Eprint {http://arxiv.org/abs/1608.02942} {arXiv:1608.02942 [hep-th]}
  \BibitemShut {NoStop}%
\bibitem [{\citenamefont {de~Rham}\ \emph
  {et~al.}(2017{\natexlab{a}})\citenamefont {de~Rham}, \citenamefont
  {Melville}, \citenamefont {Tolley},\ and\ \citenamefont
  {Zhou}}]{deRham:2017imi}%
  \BibitemOpen
  \bibfield  {author} {\bibinfo {author} {\bibfnamefont {C.}~\bibnamefont
  {de~Rham}}, \bibinfo {author} {\bibfnamefont {S.}~\bibnamefont {Melville}},
  \bibinfo {author} {\bibfnamefont {A.~J.}\ \bibnamefont {Tolley}}, \ and\
  \bibinfo {author} {\bibfnamefont {S.-Y.}\ \bibnamefont {Zhou}},\ }\href
  {\doibase 10.1007/JHEP09(2017)072} {\bibfield  {journal} {\bibinfo  {journal}
  {JHEP}\ }\textbf {\bibinfo {volume} {09}},\ \bibinfo {pages} {072} (\bibinfo
  {year} {2017}{\natexlab{a}})},\ \Eprint {http://arxiv.org/abs/1702.08577}
  {arXiv:1702.08577 [hep-th]} \BibitemShut {NoStop}%
\bibitem [{\citenamefont {Bellazzini}\ \emph {et~al.}(2018)\citenamefont
  {Bellazzini}, \citenamefont {Riva}, \citenamefont {Serra},\ and\
  \citenamefont {Sgarlata}}]{Bellazzini:2017fep}%
  \BibitemOpen
  \bibfield  {author} {\bibinfo {author} {\bibfnamefont {B.}~\bibnamefont
  {Bellazzini}}, \bibinfo {author} {\bibfnamefont {F.}~\bibnamefont {Riva}},
  \bibinfo {author} {\bibfnamefont {J.}~\bibnamefont {Serra}}, \ and\ \bibinfo
  {author} {\bibfnamefont {F.}~\bibnamefont {Sgarlata}},\ }\href {\doibase
  10.1103/PhysRevLett.120.161101} {\bibfield  {journal} {\bibinfo  {journal}
  {Phys. Rev. Lett.}\ }\textbf {\bibinfo {volume} {120}},\ \bibinfo {pages}
  {161101} (\bibinfo {year} {2018})},\ \Eprint
  {http://arxiv.org/abs/1710.02539} {arXiv:1710.02539 [hep-th]} \BibitemShut
  {NoStop}%
\bibitem [{\citenamefont {de~Rham}\ \emph {et~al.}(2018)\citenamefont
  {de~Rham}, \citenamefont {Melville},\ and\ \citenamefont
  {Tolley}}]{deRham:2017xox}%
  \BibitemOpen
  \bibfield  {author} {\bibinfo {author} {\bibfnamefont {C.}~\bibnamefont
  {de~Rham}}, \bibinfo {author} {\bibfnamefont {S.}~\bibnamefont {Melville}}, \
  and\ \bibinfo {author} {\bibfnamefont {A.~J.}\ \bibnamefont {Tolley}},\
  }\href {\doibase 10.1007/JHEP04(2018)083} {\bibfield  {journal} {\bibinfo
  {journal} {JHEP}\ }\textbf {\bibinfo {volume} {04}},\ \bibinfo {pages} {083}
  (\bibinfo {year} {2018})},\ \Eprint {http://arxiv.org/abs/1710.09611}
  {arXiv:1710.09611 [hep-th]} \BibitemShut {NoStop}%
\bibitem [{\citenamefont {de~Rham}\ \emph
  {et~al.}(2017{\natexlab{b}})\citenamefont {de~Rham}, \citenamefont
  {Melville}, \citenamefont {Tolley},\ and\ \citenamefont
  {Zhou}}]{deRham:2017avq}%
  \BibitemOpen
  \bibfield  {author} {\bibinfo {author} {\bibfnamefont {C.}~\bibnamefont
  {de~Rham}}, \bibinfo {author} {\bibfnamefont {S.}~\bibnamefont {Melville}},
  \bibinfo {author} {\bibfnamefont {A.~J.}\ \bibnamefont {Tolley}}, \ and\
  \bibinfo {author} {\bibfnamefont {S.-Y.}\ \bibnamefont {Zhou}},\ }\href
  {\doibase 10.1103/PhysRevD.96.081702} {\bibfield  {journal} {\bibinfo
  {journal} {Phys. Rev. D}\ }\textbf {\bibinfo {volume} {96}},\ \bibinfo
  {pages} {081702} (\bibinfo {year} {2017}{\natexlab{b}})},\ \Eprint
  {http://arxiv.org/abs/1702.06134} {arXiv:1702.06134 [hep-th]} \BibitemShut
  {NoStop}%
\bibitem [{\citenamefont {Bellazzini}\ \emph {et~al.}(2019)\citenamefont
  {Bellazzini}, \citenamefont {Lewandowski},\ and\ \citenamefont
  {Serra}}]{Bellazzini:2019xts}%
  \BibitemOpen
  \bibfield  {author} {\bibinfo {author} {\bibfnamefont {B.}~\bibnamefont
  {Bellazzini}}, \bibinfo {author} {\bibfnamefont {M.}~\bibnamefont
  {Lewandowski}}, \ and\ \bibinfo {author} {\bibfnamefont {J.}~\bibnamefont
  {Serra}},\ }\href {\doibase 10.1103/PhysRevLett.123.251103} {\bibfield
  {journal} {\bibinfo  {journal} {Phys. Rev. Lett.}\ }\textbf {\bibinfo
  {volume} {123}},\ \bibinfo {pages} {251103} (\bibinfo {year} {2019})},\
  \Eprint {http://arxiv.org/abs/1902.03250} {arXiv:1902.03250 [hep-th]}
  \BibitemShut {NoStop}%
\bibitem [{\citenamefont {Melville}\ and\ \citenamefont
  {Noller}(2020)}]{Melville:2019wyy}%
  \BibitemOpen
  \bibfield  {author} {\bibinfo {author} {\bibfnamefont {S.}~\bibnamefont
  {Melville}}\ and\ \bibinfo {author} {\bibfnamefont {J.}~\bibnamefont
  {Noller}},\ }\href {\doibase 10.1103/PhysRevD.101.021502} {\bibfield
  {journal} {\bibinfo  {journal} {Phys. Rev. D}\ }\textbf {\bibinfo {volume}
  {101}},\ \bibinfo {pages} {021502} (\bibinfo {year} {2020})},\ \Eprint
  {http://arxiv.org/abs/1904.05874} {arXiv:1904.05874 [astro-ph.CO]}
  \BibitemShut {NoStop}%
\bibitem [{\citenamefont {Ye}\ and\ \citenamefont {Piao}(2020)}]{Ye:2019oxx}%
  \BibitemOpen
  \bibfield  {author} {\bibinfo {author} {\bibfnamefont {G.}~\bibnamefont
  {Ye}}\ and\ \bibinfo {author} {\bibfnamefont {Y.-S.}\ \bibnamefont {Piao}},\
  }\href {\doibase 10.1140/epjc/s10052-020-7973-z} {\bibfield  {journal}
  {\bibinfo  {journal} {Eur. Phys. J. C}\ }\textbf {\bibinfo {volume} {80}},\
  \bibinfo {pages} {421} (\bibinfo {year} {2020})},\ \Eprint
  {http://arxiv.org/abs/1908.08644} {arXiv:1908.08644 [hep-th]} \BibitemShut
  {NoStop}%
\bibitem [{\citenamefont {Tokuda}\ \emph {et~al.}(2020)\citenamefont {Tokuda},
  \citenamefont {Aoki},\ and\ \citenamefont {Hirano}}]{Tokuda:2020mlf}%
  \BibitemOpen
  \bibfield  {author} {\bibinfo {author} {\bibfnamefont {J.}~\bibnamefont
  {Tokuda}}, \bibinfo {author} {\bibfnamefont {K.}~\bibnamefont {Aoki}}, \ and\
  \bibinfo {author} {\bibfnamefont {S.}~\bibnamefont {Hirano}},\ }\href
  {\doibase 10.1007/JHEP11(2020)054} {\bibfield  {journal} {\bibinfo  {journal}
  {JHEP}\ }\textbf {\bibinfo {volume} {11}},\ \bibinfo {pages} {054} (\bibinfo
  {year} {2020})},\ \Eprint {http://arxiv.org/abs/2007.15009} {arXiv:2007.15009
  [hep-th]} \BibitemShut {NoStop}%
\bibitem [{\citenamefont {Kennedy}\ and\ \citenamefont
  {Lombriser}(2020)}]{Kennedy:2020ehn}%
  \BibitemOpen
  \bibfield  {author} {\bibinfo {author} {\bibfnamefont {J.}~\bibnamefont
  {Kennedy}}\ and\ \bibinfo {author} {\bibfnamefont {L.}~\bibnamefont
  {Lombriser}},\ }\href {\doibase 10.1103/PhysRevD.102.044062} {\bibfield
  {journal} {\bibinfo  {journal} {Phys. Rev. D}\ }\textbf {\bibinfo {volume}
  {102}},\ \bibinfo {pages} {044062} (\bibinfo {year} {2020})},\ \Eprint
  {http://arxiv.org/abs/2003.05318} {arXiv:2003.05318 [gr-qc]} \BibitemShut
  {NoStop}%
\bibitem [{\citenamefont {de~Rham}\ and\ \citenamefont
  {Tolley}(2020)}]{deRham:2020zyh}%
  \BibitemOpen
  \bibfield  {author} {\bibinfo {author} {\bibfnamefont {C.}~\bibnamefont
  {de~Rham}}\ and\ \bibinfo {author} {\bibfnamefont {A.~J.}\ \bibnamefont
  {Tolley}},\ }\href {\doibase 10.1103/PhysRevD.102.084048} {\bibfield
  {journal} {\bibinfo  {journal} {Phys. Rev. D}\ }\textbf {\bibinfo {volume}
  {102}},\ \bibinfo {pages} {084048} (\bibinfo {year} {2020})},\ \Eprint
  {http://arxiv.org/abs/2007.01847} {arXiv:2007.01847 [hep-th]} \BibitemShut
  {NoStop}%
\bibitem [{\citenamefont {de~Rham}\ \emph {et~al.}(2021)\citenamefont
  {de~Rham}, \citenamefont {Melville},\ and\ \citenamefont
  {Noller}}]{deRham:2021fpu}%
  \BibitemOpen
  \bibfield  {author} {\bibinfo {author} {\bibfnamefont {C.}~\bibnamefont
  {de~Rham}}, \bibinfo {author} {\bibfnamefont {S.}~\bibnamefont {Melville}}, \
  and\ \bibinfo {author} {\bibfnamefont {J.}~\bibnamefont {Noller}},\ }\href
  {\doibase 10.1088/1475-7516/2021/08/018} {\bibfield  {journal} {\bibinfo
  {journal} {JCAP}\ }\textbf {\bibinfo {volume} {08}},\ \bibinfo {pages} {018}
  (\bibinfo {year} {2021})},\ \Eprint {http://arxiv.org/abs/2103.06855}
  {arXiv:2103.06855 [astro-ph.CO]} \BibitemShut {NoStop}%
\bibitem [{\citenamefont {Traykova}\ \emph {et~al.}(2021)\citenamefont
  {Traykova}, \citenamefont {Bellini}, \citenamefont {Ferreira}, \citenamefont
  {Garc\'\i{}a-Garc\'\i{}a}, \citenamefont {Noller},\ and\ \citenamefont
  {Zumalac\'arregui}}]{Traykova:2021hbr}%
  \BibitemOpen
  \bibfield  {author} {\bibinfo {author} {\bibfnamefont {D.}~\bibnamefont
  {Traykova}}, \bibinfo {author} {\bibfnamefont {E.}~\bibnamefont {Bellini}},
  \bibinfo {author} {\bibfnamefont {P.~G.}\ \bibnamefont {Ferreira}}, \bibinfo
  {author} {\bibfnamefont {C.}~\bibnamefont {Garc\'\i{}a-Garc\'\i{}a}},
  \bibinfo {author} {\bibfnamefont {J.}~\bibnamefont {Noller}}, \ and\ \bibinfo
  {author} {\bibfnamefont {M.}~\bibnamefont {Zumalac\'arregui}},\ }\href
  {\doibase 10.1103/PhysRevD.104.083502} {\bibfield  {journal} {\bibinfo
  {journal} {Phys. Rev. D}\ }\textbf {\bibinfo {volume} {104}},\ \bibinfo
  {pages} {083502} (\bibinfo {year} {2021})},\ \Eprint
  {http://arxiv.org/abs/2103.11195} {arXiv:2103.11195 [astro-ph.CO]}
  \BibitemShut {NoStop}%
\bibitem [{\citenamefont {Grall}\ and\ \citenamefont
  {Melville}(2022)}]{Grall:2021xxm}%
  \BibitemOpen
  \bibfield  {author} {\bibinfo {author} {\bibfnamefont {T.}~\bibnamefont
  {Grall}}\ and\ \bibinfo {author} {\bibfnamefont {S.}~\bibnamefont
  {Melville}},\ }\href {\doibase 10.1103/PhysRevD.105.L121301} {\bibfield
  {journal} {\bibinfo  {journal} {Phys. Rev. D}\ }\textbf {\bibinfo {volume}
  {105}},\ \bibinfo {pages} {L121301} (\bibinfo {year} {2022})},\ \Eprint
  {http://arxiv.org/abs/2102.05683} {arXiv:2102.05683 [hep-th]} \BibitemShut
  {NoStop}%
\bibitem [{\citenamefont {Melville}\ and\ \citenamefont
  {Noller}(2022)}]{Melville:2022ykg}%
  \BibitemOpen
  \bibfield  {author} {\bibinfo {author} {\bibfnamefont {S.}~\bibnamefont
  {Melville}}\ and\ \bibinfo {author} {\bibfnamefont {J.}~\bibnamefont
  {Noller}},\ }\href {\doibase 10.1088/1475-7516/2022/06/031} {\bibfield
  {journal} {\bibinfo  {journal} {JCAP}\ }\textbf {\bibinfo {volume} {06}},\
  \bibinfo {pages} {031} (\bibinfo {year} {2022})},\ \Eprint
  {http://arxiv.org/abs/2202.01222} {arXiv:2202.01222 [hep-th]} \BibitemShut
  {NoStop}%
\bibitem [{\citenamefont {Xu}\ and\ \citenamefont {Zhou}(2023)}]{Xu:2023lpq}%
  \BibitemOpen
  \bibfield  {author} {\bibinfo {author} {\bibfnamefont {H.}~\bibnamefont
  {Xu}}\ and\ \bibinfo {author} {\bibfnamefont {S.-Y.}\ \bibnamefont {Zhou}},\
  }\href {\doibase 10.1088/1475-7516/2023/11/076} {\bibfield  {journal}
  {\bibinfo  {journal} {JCAP}\ }\textbf {\bibinfo {volume} {11}},\ \bibinfo
  {pages} {076} (\bibinfo {year} {2023})},\ \Eprint
  {http://arxiv.org/abs/2306.06639} {arXiv:2306.06639 [hep-th]} \BibitemShut
  {NoStop}%
\bibitem [{\citenamefont {Bellazzini}\ \emph {et~al.}(2024)\citenamefont
  {Bellazzini}, \citenamefont {Isabella}, \citenamefont {Ricossa},\ and\
  \citenamefont {Riva}}]{Bellazzini:2023nqj}%
  \BibitemOpen
  \bibfield  {author} {\bibinfo {author} {\bibfnamefont {B.}~\bibnamefont
  {Bellazzini}}, \bibinfo {author} {\bibfnamefont {G.}~\bibnamefont
  {Isabella}}, \bibinfo {author} {\bibfnamefont {S.}~\bibnamefont {Ricossa}}, \
  and\ \bibinfo {author} {\bibfnamefont {F.}~\bibnamefont {Riva}},\ }\href
  {\doibase 10.1103/PhysRevD.109.024051} {\bibfield  {journal} {\bibinfo
  {journal} {Phys. Rev. D}\ }\textbf {\bibinfo {volume} {109}},\ \bibinfo
  {pages} {024051} (\bibinfo {year} {2024})},\ \Eprint
  {http://arxiv.org/abs/2304.02550} {arXiv:2304.02550 [hep-th]} \BibitemShut
  {NoStop}%
\bibitem [{\citenamefont {Raveri}\ \emph {et~al.}(2017)\citenamefont {Raveri},
  \citenamefont {Bull}, \citenamefont {Silvestri},\ and\ \citenamefont
  {Pogosian}}]{Raveri:2017qvt}%
  \BibitemOpen
  \bibfield  {author} {\bibinfo {author} {\bibfnamefont {M.}~\bibnamefont
  {Raveri}}, \bibinfo {author} {\bibfnamefont {P.}~\bibnamefont {Bull}},
  \bibinfo {author} {\bibfnamefont {A.}~\bibnamefont {Silvestri}}, \ and\
  \bibinfo {author} {\bibfnamefont {L.}~\bibnamefont {Pogosian}},\ }\href
  {\doibase 10.1103/PhysRevD.96.083509} {\bibfield  {journal} {\bibinfo
  {journal} {Phys. Rev.}\ }\textbf {\bibinfo {volume} {D96}},\ \bibinfo {pages}
  {083509} (\bibinfo {year} {2017})},\ \Eprint
  {http://arxiv.org/abs/1703.05297} {arXiv:1703.05297 [astro-ph.CO]}
  \BibitemShut {NoStop}%
\bibitem [{\citenamefont {Peirone}\ \emph {et~al.}(2018)\citenamefont
  {Peirone}, \citenamefont {Koyama}, \citenamefont {Pogosian}, \citenamefont
  {Raveri},\ and\ \citenamefont {Silvestri}}]{Peirone:2017ywi}%
  \BibitemOpen
  \bibfield  {author} {\bibinfo {author} {\bibfnamefont {S.}~\bibnamefont
  {Peirone}}, \bibinfo {author} {\bibfnamefont {K.}~\bibnamefont {Koyama}},
  \bibinfo {author} {\bibfnamefont {L.}~\bibnamefont {Pogosian}}, \bibinfo
  {author} {\bibfnamefont {M.}~\bibnamefont {Raveri}}, \ and\ \bibinfo {author}
  {\bibfnamefont {A.}~\bibnamefont {Silvestri}},\ }\href {\doibase
  10.1103/PhysRevD.97.043519} {\bibfield  {journal} {\bibinfo  {journal} {Phys.
  Rev.}\ }\textbf {\bibinfo {volume} {D97}},\ \bibinfo {pages} {043519}
  (\bibinfo {year} {2018})},\ \Eprint {http://arxiv.org/abs/1712.00444}
  {arXiv:1712.00444 [astro-ph.CO]} \BibitemShut {NoStop}%
\bibitem [{\citenamefont {Espejo}\ \emph {et~al.}(2019)\citenamefont {Espejo},
  \citenamefont {Peirone}, \citenamefont {Raveri}, \citenamefont {Koyama},
  \citenamefont {Pogosian},\ and\ \citenamefont {Silvestri}}]{Espejo:2018hxa}%
  \BibitemOpen
  \bibfield  {author} {\bibinfo {author} {\bibfnamefont {J.}~\bibnamefont
  {Espejo}}, \bibinfo {author} {\bibfnamefont {S.}~\bibnamefont {Peirone}},
  \bibinfo {author} {\bibfnamefont {M.}~\bibnamefont {Raveri}}, \bibinfo
  {author} {\bibfnamefont {K.}~\bibnamefont {Koyama}}, \bibinfo {author}
  {\bibfnamefont {L.}~\bibnamefont {Pogosian}}, \ and\ \bibinfo {author}
  {\bibfnamefont {A.}~\bibnamefont {Silvestri}},\ }\href {\doibase
  10.1103/PhysRevD.99.023512} {\bibfield  {journal} {\bibinfo  {journal} {Phys.
  Rev.}\ }\textbf {\bibinfo {volume} {D99}},\ \bibinfo {pages} {023512}
  (\bibinfo {year} {2019})},\ \Eprint {http://arxiv.org/abs/1809.01121}
  {arXiv:1809.01121 [astro-ph.CO]} \BibitemShut {NoStop}%
\bibitem [{\citenamefont {Frusciante}\ \emph {et~al.}(2019)\citenamefont
  {Frusciante}, \citenamefont {Papadomanolakis}, \citenamefont {Peirone},\ and\
  \citenamefont {Silvestri}}]{Frusciante:2018vht}%
  \BibitemOpen
  \bibfield  {author} {\bibinfo {author} {\bibfnamefont {N.}~\bibnamefont
  {Frusciante}}, \bibinfo {author} {\bibfnamefont {G.}~\bibnamefont
  {Papadomanolakis}}, \bibinfo {author} {\bibfnamefont {S.}~\bibnamefont
  {Peirone}}, \ and\ \bibinfo {author} {\bibfnamefont {A.}~\bibnamefont
  {Silvestri}},\ }\href {\doibase 10.1088/1475-7516/2019/02/029} {\bibfield
  {journal} {\bibinfo  {journal} {JCAP}\ }\textbf {\bibinfo {volume} {1902}},\
  \bibinfo {pages} {029} (\bibinfo {year} {2019})},\ \Eprint
  {http://arxiv.org/abs/1810.03461} {arXiv:1810.03461 [gr-qc]} \BibitemShut
  {NoStop}%
\bibitem [{\citenamefont {Frusciante}\ \emph {et~al.}(2016)\citenamefont
  {Frusciante}, \citenamefont {Papadomanolakis},\ and\ \citenamefont
  {Silvestri}}]{Frusciante:2016xoj}%
  \BibitemOpen
  \bibfield  {author} {\bibinfo {author} {\bibfnamefont {N.}~\bibnamefont
  {Frusciante}}, \bibinfo {author} {\bibfnamefont {G.}~\bibnamefont
  {Papadomanolakis}}, \ and\ \bibinfo {author} {\bibfnamefont {A.}~\bibnamefont
  {Silvestri}},\ }\href {\doibase 10.1088/1475-7516/2016/07/018} {\bibfield
  {journal} {\bibinfo  {journal} {JCAP}\ }\textbf {\bibinfo {volume} {07}},\
  \bibinfo {pages} {018} (\bibinfo {year} {2016})},\ \Eprint
  {http://arxiv.org/abs/1601.04064} {arXiv:1601.04064 [gr-qc]} \BibitemShut
  {NoStop}%
\bibitem [{\citenamefont {De~Felice}\ \emph {et~al.}(2017)\citenamefont
  {De~Felice}, \citenamefont {Frusciante},\ and\ \citenamefont
  {Papadomanolakis}}]{DeFelice:2016ucp}%
  \BibitemOpen
  \bibfield  {author} {\bibinfo {author} {\bibfnamefont {A.}~\bibnamefont
  {De~Felice}}, \bibinfo {author} {\bibfnamefont {N.}~\bibnamefont
  {Frusciante}}, \ and\ \bibinfo {author} {\bibfnamefont {G.}~\bibnamefont
  {Papadomanolakis}},\ }\href {\doibase 10.1088/1475-7516/2017/03/027}
  {\bibfield  {journal} {\bibinfo  {journal} {JCAP}\ }\textbf {\bibinfo
  {volume} {03}},\ \bibinfo {pages} {027} (\bibinfo {year} {2017})},\ \Eprint
  {http://arxiv.org/abs/1609.03599} {arXiv:1609.03599 [gr-qc]} \BibitemShut
  {NoStop}%
\bibitem [{\citenamefont {Hu}\ \emph {et~al.}(2014)\citenamefont {Hu},
  \citenamefont {Raveri}, \citenamefont {Frusciante},\ and\ \citenamefont
  {Silvestri}}]{Hu:2013twa}%
  \BibitemOpen
  \bibfield  {author} {\bibinfo {author} {\bibfnamefont {B.}~\bibnamefont
  {Hu}}, \bibinfo {author} {\bibfnamefont {M.}~\bibnamefont {Raveri}}, \bibinfo
  {author} {\bibfnamefont {N.}~\bibnamefont {Frusciante}}, \ and\ \bibinfo
  {author} {\bibfnamefont {A.}~\bibnamefont {Silvestri}},\ }\href {\doibase
  10.1103/PhysRevD.89.103530} {\bibfield  {journal} {\bibinfo  {journal} {Phys.
  Rev.}\ }\textbf {\bibinfo {volume} {D89}},\ \bibinfo {pages} {103530}
  (\bibinfo {year} {2014})},\ \Eprint {http://arxiv.org/abs/1312.5742}
  {arXiv:1312.5742 [astro-ph.CO]} \BibitemShut {NoStop}%
\bibitem [{\citenamefont {Raveri}\ \emph {et~al.}(2014)\citenamefont {Raveri},
  \citenamefont {Hu}, \citenamefont {Frusciante},\ and\ \citenamefont
  {Silvestri}}]{Raveri:2014cka}%
  \BibitemOpen
  \bibfield  {author} {\bibinfo {author} {\bibfnamefont {M.}~\bibnamefont
  {Raveri}}, \bibinfo {author} {\bibfnamefont {B.}~\bibnamefont {Hu}}, \bibinfo
  {author} {\bibfnamefont {N.}~\bibnamefont {Frusciante}}, \ and\ \bibinfo
  {author} {\bibfnamefont {A.}~\bibnamefont {Silvestri}},\ }\href {\doibase
  10.1103/PhysRevD.90.043513} {\bibfield  {journal} {\bibinfo  {journal} {Phys.
  Rev.}\ }\textbf {\bibinfo {volume} {D90}},\ \bibinfo {pages} {043513}
  (\bibinfo {year} {2014})},\ \Eprint {http://arxiv.org/abs/1405.1022}
  {arXiv:1405.1022 [astro-ph.CO]} \BibitemShut {NoStop}%
\bibitem [{\citenamefont {Lewis}\ \emph {et~al.}(2000)\citenamefont {Lewis},
  \citenamefont {Challinor},\ and\ \citenamefont {Lasenby}}]{Lewis:1999bs}%
  \BibitemOpen
  \bibfield  {author} {\bibinfo {author} {\bibfnamefont {A.}~\bibnamefont
  {Lewis}}, \bibinfo {author} {\bibfnamefont {A.}~\bibnamefont {Challinor}}, \
  and\ \bibinfo {author} {\bibfnamefont {A.}~\bibnamefont {Lasenby}},\ }\href
  {\doibase 10.1086/309179} {\bibfield  {journal} {\bibinfo  {journal}
  {Astrophys. J.}\ }\textbf {\bibinfo {volume} {538}},\ \bibinfo {pages} {473}
  (\bibinfo {year} {2000})},\ \Eprint {http://arxiv.org/abs/astro-ph/9911177}
  {arXiv:astro-ph/9911177 [astro-ph]} \BibitemShut {NoStop}%
\bibitem [{cam()}]{camb}%
  \BibitemOpen
  \href@noop {} {}\bibinfo {howpublished} {\url{http://camb.info}}\BibitemShut
  {NoStop}%
\bibitem [{\citenamefont {Horndeski}(1974)}]{Horndeski:1974wa}%
  \BibitemOpen
  \bibfield  {author} {\bibinfo {author} {\bibfnamefont {G.~W.}\ \bibnamefont
  {Horndeski}},\ }\href {\doibase 10.1007/BF01807638} {\bibfield  {journal}
  {\bibinfo  {journal} {Int. J. Theor. Phys.}\ }\textbf {\bibinfo {volume}
  {10}},\ \bibinfo {pages} {363} (\bibinfo {year} {1974})}\BibitemShut
  {NoStop}%
\bibitem [{\citenamefont {Horndeski}\ and\ \citenamefont
  {Silvestri}(2024)}]{Horndeski:2024sjk}%
  \BibitemOpen
  \bibfield  {author} {\bibinfo {author} {\bibfnamefont {G.~W.}\ \bibnamefont
  {Horndeski}}\ and\ \bibinfo {author} {\bibfnamefont {A.}~\bibnamefont
  {Silvestri}},\ }\href {\doibase 10.1007/s10773-024-05558-2} {\bibfield
  {journal} {\bibinfo  {journal} {Int. J. Theor. Phys.}\ }\textbf {\bibinfo
  {volume} {63}},\ \bibinfo {pages} {38} (\bibinfo {year} {2024})},\ \Eprint
  {http://arxiv.org/abs/2402.07538} {arXiv:2402.07538 [gr-qc]} \BibitemShut
  {NoStop}%
\bibitem [{euc()}]{euclid}%
  \BibitemOpen
  \href@noop {} {\emph {\bibinfo {title}
  {\url{http://www.euclid-ec.org}}}}\BibitemShut {NoStop}%
\bibitem [{lss()}]{lsst}%
  \BibitemOpen
  \href@noop {} {\emph {\bibinfo {title}
  {\url{http://www.lsst.org}}}}\BibitemShut {NoStop}%
\bibitem [{\citenamefont {Kobayashi}\ \emph {et~al.}(2011)\citenamefont
  {Kobayashi}, \citenamefont {Yamaguchi},\ and\ \citenamefont
  {Yokoyama}}]{Kobayashi:2011nu}%
  \BibitemOpen
  \bibfield  {author} {\bibinfo {author} {\bibfnamefont {T.}~\bibnamefont
  {Kobayashi}}, \bibinfo {author} {\bibfnamefont {M.}~\bibnamefont
  {Yamaguchi}}, \ and\ \bibinfo {author} {\bibfnamefont {J.}~\bibnamefont
  {Yokoyama}},\ }\href {\doibase 10.1143/PTP.126.511} {\bibfield  {journal}
  {\bibinfo  {journal} {Prog. Theor. Phys.}\ }\textbf {\bibinfo {volume}
  {126}},\ \bibinfo {pages} {511} (\bibinfo {year} {2011})},\ \Eprint
  {http://arxiv.org/abs/1105.5723} {arXiv:1105.5723 [hep-th]} \BibitemShut
  {NoStop}%
\bibitem [{\citenamefont {Nicolis}\ \emph {et~al.}(2009)\citenamefont
  {Nicolis}, \citenamefont {Rattazzi},\ and\ \citenamefont
  {Trincherini}}]{Nicolis:2008in}%
  \BibitemOpen
  \bibfield  {author} {\bibinfo {author} {\bibfnamefont {A.}~\bibnamefont
  {Nicolis}}, \bibinfo {author} {\bibfnamefont {R.}~\bibnamefont {Rattazzi}}, \
  and\ \bibinfo {author} {\bibfnamefont {E.}~\bibnamefont {Trincherini}},\
  }\href {\doibase 10.1103/PhysRevD.79.064036} {\bibfield  {journal} {\bibinfo
  {journal} {Phys. Rev.}\ }\textbf {\bibinfo {volume} {D79}},\ \bibinfo {pages}
  {064036} (\bibinfo {year} {2009})},\ \Eprint {http://arxiv.org/abs/0811.2197}
  {arXiv:0811.2197 [hep-th]} \BibitemShut {NoStop}%
\bibitem [{\citenamefont {De~Felice}\ and\ \citenamefont
  {Tsujikawa}(2011)}]{DeFelice:2010nf}%
  \BibitemOpen
  \bibfield  {author} {\bibinfo {author} {\bibfnamefont {A.}~\bibnamefont
  {De~Felice}}\ and\ \bibinfo {author} {\bibfnamefont {S.}~\bibnamefont
  {Tsujikawa}},\ }\href {\doibase 10.1103/PhysRevD.84.124029} {\bibfield
  {journal} {\bibinfo  {journal} {Phys. Rev. D}\ }\textbf {\bibinfo {volume}
  {84}},\ \bibinfo {pages} {124029} (\bibinfo {year} {2011})},\ \Eprint
  {http://arxiv.org/abs/1008.4236} {arXiv:1008.4236 [hep-th]} \BibitemShut
  {NoStop}%
\bibitem [{\citenamefont {Hohmann}(2015)}]{Hohmann:2015kra}%
  \BibitemOpen
  \bibfield  {author} {\bibinfo {author} {\bibfnamefont {M.}~\bibnamefont
  {Hohmann}},\ }\href {\doibase 10.1103/PhysRevD.92.064019} {\bibfield
  {journal} {\bibinfo  {journal} {Phys. Rev. D}\ }\textbf {\bibinfo {volume}
  {92}},\ \bibinfo {pages} {064019} (\bibinfo {year} {2015})},\ \Eprint
  {http://arxiv.org/abs/1506.04253} {arXiv:1506.04253 [gr-qc]} \BibitemShut
  {NoStop}%
\bibitem [{\citenamefont {Bloomfield}(2013)}]{Bloomfield:2013efa}%
  \BibitemOpen
  \bibfield  {author} {\bibinfo {author} {\bibfnamefont {J.}~\bibnamefont
  {Bloomfield}},\ }\href {\doibase 10.1088/1475-7516/2013/12/044} {\bibfield
  {journal} {\bibinfo  {journal} {JCAP}\ }\textbf {\bibinfo {volume} {1312}},\
  \bibinfo {pages} {044} (\bibinfo {year} {2013})},\ \Eprint
  {http://arxiv.org/abs/1304.6712} {arXiv:1304.6712 [astro-ph.CO]} \BibitemShut
  {NoStop}%
\bibitem [{\citenamefont {Gleyzes}\ \emph {et~al.}(2013)\citenamefont
  {Gleyzes}, \citenamefont {Langlois}, \citenamefont {Piazza},\ and\
  \citenamefont {Vernizzi}}]{Gleyzes:2013ooa}%
  \BibitemOpen
  \bibfield  {author} {\bibinfo {author} {\bibfnamefont {J.}~\bibnamefont
  {Gleyzes}}, \bibinfo {author} {\bibfnamefont {D.}~\bibnamefont {Langlois}},
  \bibinfo {author} {\bibfnamefont {F.}~\bibnamefont {Piazza}}, \ and\ \bibinfo
  {author} {\bibfnamefont {F.}~\bibnamefont {Vernizzi}},\ }\href {\doibase
  10.1088/1475-7516/2013/08/025} {\bibfield  {journal} {\bibinfo  {journal}
  {JCAP}\ }\textbf {\bibinfo {volume} {1308}},\ \bibinfo {pages} {025}
  (\bibinfo {year} {2013})},\ \Eprint {http://arxiv.org/abs/1304.4840}
  {arXiv:1304.4840 [hep-th]} \BibitemShut {NoStop}%
\bibitem [{\citenamefont {Kennedy}\ \emph {et~al.}(2017)\citenamefont
  {Kennedy}, \citenamefont {Lombriser},\ and\ \citenamefont
  {Taylor}}]{Kennedy:2017sof}%
  \BibitemOpen
  \bibfield  {author} {\bibinfo {author} {\bibfnamefont {J.}~\bibnamefont
  {Kennedy}}, \bibinfo {author} {\bibfnamefont {L.}~\bibnamefont {Lombriser}},
  \ and\ \bibinfo {author} {\bibfnamefont {A.}~\bibnamefont {Taylor}},\ }\href
  {\doibase 10.1103/PhysRevD.96.084051} {\bibfield  {journal} {\bibinfo
  {journal} {Phys. Rev.}\ }\textbf {\bibinfo {volume} {D96}},\ \bibinfo {pages}
  {084051} (\bibinfo {year} {2017})},\ \Eprint
  {http://arxiv.org/abs/1705.09290} {arXiv:1705.09290 [gr-qc]} \BibitemShut
  {NoStop}%
\bibitem [{\citenamefont {Zucca}\ \emph {et~al.}(2019)\citenamefont {Zucca},
  \citenamefont {Pogosian}, \citenamefont {Silvestri},\ and\ \citenamefont
  {Zhao}}]{Zucca:2019xhg}%
  \BibitemOpen
  \bibfield  {author} {\bibinfo {author} {\bibfnamefont {A.}~\bibnamefont
  {Zucca}}, \bibinfo {author} {\bibfnamefont {L.}~\bibnamefont {Pogosian}},
  \bibinfo {author} {\bibfnamefont {A.}~\bibnamefont {Silvestri}}, \ and\
  \bibinfo {author} {\bibfnamefont {G.-B.}\ \bibnamefont {Zhao}},\ }\href
  {\doibase 10.1088/1475-7516/2019/05/001} {\bibfield  {journal} {\bibinfo
  {journal} {JCAP}\ }\textbf {\bibinfo {volume} {05}},\ \bibinfo {pages} {001}
  (\bibinfo {year} {2019})},\ \Eprint {http://arxiv.org/abs/1901.05956}
  {arXiv:1901.05956 [astro-ph.CO]} \BibitemShut {NoStop}%
\bibitem [{\citenamefont {Pogosian}\ \emph {et~al.}(2010)\citenamefont
  {Pogosian}, \citenamefont {Silvestri}, \citenamefont {Koyama},\ and\
  \citenamefont {Zhao}}]{Pogosian:2010tj}%
  \BibitemOpen
  \bibfield  {author} {\bibinfo {author} {\bibfnamefont {L.}~\bibnamefont
  {Pogosian}}, \bibinfo {author} {\bibfnamefont {A.}~\bibnamefont {Silvestri}},
  \bibinfo {author} {\bibfnamefont {K.}~\bibnamefont {Koyama}}, \ and\ \bibinfo
  {author} {\bibfnamefont {G.-B.}\ \bibnamefont {Zhao}},\ }\href {\doibase
  10.1103/PhysRevD.81.104023} {\bibfield  {journal} {\bibinfo  {journal} {Phys.
  Rev. D}\ }\textbf {\bibinfo {volume} {81}},\ \bibinfo {pages} {104023}
  (\bibinfo {year} {2010})},\ \Eprint {http://arxiv.org/abs/1002.2382}
  {arXiv:1002.2382 [astro-ph.CO]} \BibitemShut {NoStop}%
\bibitem [{\citenamefont {Amendola}\ \emph {et~al.}(2008)\citenamefont
  {Amendola}, \citenamefont {Kunz},\ and\ \citenamefont
  {Sapone}}]{Amendola:2007rr}%
  \BibitemOpen
  \bibfield  {author} {\bibinfo {author} {\bibfnamefont {L.}~\bibnamefont
  {Amendola}}, \bibinfo {author} {\bibfnamefont {M.}~\bibnamefont {Kunz}}, \
  and\ \bibinfo {author} {\bibfnamefont {D.}~\bibnamefont {Sapone}},\ }\href
  {\doibase 10.1088/1475-7516/2008/04/013} {\bibfield  {journal} {\bibinfo
  {journal} {JCAP}\ }\textbf {\bibinfo {volume} {04}},\ \bibinfo {pages} {013}
  (\bibinfo {year} {2008})},\ \Eprint {http://arxiv.org/abs/0704.2421}
  {arXiv:0704.2421 [astro-ph]} \BibitemShut {NoStop}%
\bibitem [{\citenamefont {Jain}\ and\ \citenamefont
  {Zhang}(2008)}]{Jain:2007yk}%
  \BibitemOpen
  \bibfield  {author} {\bibinfo {author} {\bibfnamefont {B.}~\bibnamefont
  {Jain}}\ and\ \bibinfo {author} {\bibfnamefont {P.}~\bibnamefont {Zhang}},\
  }\href {\doibase 10.1103/PhysRevD.78.063503} {\bibfield  {journal} {\bibinfo
  {journal} {Phys. Rev. D}\ }\textbf {\bibinfo {volume} {78}},\ \bibinfo
  {pages} {063503} (\bibinfo {year} {2008})},\ \Eprint
  {http://arxiv.org/abs/0709.2375} {arXiv:0709.2375 [astro-ph]} \BibitemShut
  {NoStop}%
\bibitem [{\citenamefont {Bertschinger}\ and\ \citenamefont
  {Zukin}(2008)}]{Bertschinger:2008zb}%
  \BibitemOpen
  \bibfield  {author} {\bibinfo {author} {\bibfnamefont {E.}~\bibnamefont
  {Bertschinger}}\ and\ \bibinfo {author} {\bibfnamefont {P.}~\bibnamefont
  {Zukin}},\ }\href {\doibase 10.1103/PhysRevD.78.024015} {\bibfield  {journal}
  {\bibinfo  {journal} {Phys. Rev. D}\ }\textbf {\bibinfo {volume} {78}},\
  \bibinfo {pages} {024015} (\bibinfo {year} {2008})},\ \Eprint
  {http://arxiv.org/abs/0801.2431} {arXiv:0801.2431 [astro-ph]} \BibitemShut
  {NoStop}%
\bibitem [{\citenamefont {Song}\ \emph {et~al.}(2011)\citenamefont {Song},
  \citenamefont {Zhao}, \citenamefont {Bacon}, \citenamefont {Koyama},
  \citenamefont {Nichol},\ and\ \citenamefont {Pogosian}}]{Song:2010fg}%
  \BibitemOpen
  \bibfield  {author} {\bibinfo {author} {\bibfnamefont {Y.-S.}\ \bibnamefont
  {Song}}, \bibinfo {author} {\bibfnamefont {G.-B.}\ \bibnamefont {Zhao}},
  \bibinfo {author} {\bibfnamefont {D.}~\bibnamefont {Bacon}}, \bibinfo
  {author} {\bibfnamefont {K.}~\bibnamefont {Koyama}}, \bibinfo {author}
  {\bibfnamefont {R.~C.}\ \bibnamefont {Nichol}}, \ and\ \bibinfo {author}
  {\bibfnamefont {L.}~\bibnamefont {Pogosian}},\ }\href {\doibase
  10.1103/PhysRevD.84.083523} {\bibfield  {journal} {\bibinfo  {journal} {Phys.
  Rev. D}\ }\textbf {\bibinfo {volume} {84}},\ \bibinfo {pages} {083523}
  (\bibinfo {year} {2011})},\ \Eprint {http://arxiv.org/abs/1011.2106}
  {arXiv:1011.2106 [astro-ph.CO]} \BibitemShut {NoStop}%
\bibitem [{\citenamefont {Aghanim}\ \emph
  {et~al.}(2020{\natexlab{a}})\citenamefont {Aghanim} \emph
  {et~al.}}]{Planck:2018vyg}%
  \BibitemOpen
  \bibfield  {author} {\bibinfo {author} {\bibfnamefont {N.}~\bibnamefont
  {Aghanim}} \emph {et~al.} (\bibinfo {collaboration} {Planck}),\ }\href
  {\doibase 10.1051/0004-6361/201833910} {\bibfield  {journal} {\bibinfo
  {journal} {Astron. Astrophys.}\ }\textbf {\bibinfo {volume} {641}},\ \bibinfo
  {pages} {A6} (\bibinfo {year} {2020}{\natexlab{a}})},\ \bibinfo {note}
  {[Erratum: Astron.Astrophys. 652, C4 (2021)]},\ \Eprint
  {http://arxiv.org/abs/1807.06209} {arXiv:1807.06209 [astro-ph.CO]}
  \BibitemShut {NoStop}%
\bibitem [{\citenamefont {Pogosian}\ \emph {et~al.}(2022)\citenamefont
  {Pogosian}, \citenamefont {Raveri}, \citenamefont {Koyama}, \citenamefont
  {Martinelli}, \citenamefont {Silvestri}, \citenamefont {Zhao}, \citenamefont
  {Li}, \citenamefont {Peirone},\ and\ \citenamefont
  {Zucca}}]{Pogosian:2021mcs}%
  \BibitemOpen
  \bibfield  {author} {\bibinfo {author} {\bibfnamefont {L.}~\bibnamefont
  {Pogosian}}, \bibinfo {author} {\bibfnamefont {M.}~\bibnamefont {Raveri}},
  \bibinfo {author} {\bibfnamefont {K.}~\bibnamefont {Koyama}}, \bibinfo
  {author} {\bibfnamefont {M.}~\bibnamefont {Martinelli}}, \bibinfo {author}
  {\bibfnamefont {A.}~\bibnamefont {Silvestri}}, \bibinfo {author}
  {\bibfnamefont {G.-B.}\ \bibnamefont {Zhao}}, \bibinfo {author}
  {\bibfnamefont {J.}~\bibnamefont {Li}}, \bibinfo {author} {\bibfnamefont
  {S.}~\bibnamefont {Peirone}}, \ and\ \bibinfo {author} {\bibfnamefont
  {A.}~\bibnamefont {Zucca}},\ }\href {\doibase 10.1038/s41550-022-01808-7}
  {\bibfield  {journal} {\bibinfo  {journal} {Nature Astron.}\ }\textbf
  {\bibinfo {volume} {6}},\ \bibinfo {pages} {1484} (\bibinfo {year} {2022})},\
  \Eprint {http://arxiv.org/abs/2107.12992} {arXiv:2107.12992 [astro-ph.CO]}
  \BibitemShut {NoStop}%
\bibitem [{\citenamefont {Raveri}\ \emph {et~al.}(2023)\citenamefont {Raveri},
  \citenamefont {Pogosian}, \citenamefont {Martinelli}, \citenamefont {Koyama},
  \citenamefont {Silvestri},\ and\ \citenamefont {Zhao}}]{Raveri:2021dbu}%
  \BibitemOpen
  \bibfield  {author} {\bibinfo {author} {\bibfnamefont {M.}~\bibnamefont
  {Raveri}}, \bibinfo {author} {\bibfnamefont {L.}~\bibnamefont {Pogosian}},
  \bibinfo {author} {\bibfnamefont {M.}~\bibnamefont {Martinelli}}, \bibinfo
  {author} {\bibfnamefont {K.}~\bibnamefont {Koyama}}, \bibinfo {author}
  {\bibfnamefont {A.}~\bibnamefont {Silvestri}}, \ and\ \bibinfo {author}
  {\bibfnamefont {G.-B.}\ \bibnamefont {Zhao}},\ }\href {\doibase
  10.1088/1475-7516/2023/02/061} {\bibfield  {journal} {\bibinfo  {journal}
  {JCAP}\ }\textbf {\bibinfo {volume} {02}},\ \bibinfo {pages} {061} (\bibinfo
  {year} {2023})},\ \Eprint {http://arxiv.org/abs/2107.12990} {arXiv:2107.12990
  [astro-ph.CO]} \BibitemShut {NoStop}%
\bibitem [{\citenamefont {Abbott}\ \emph {et~al.}(2023)\citenamefont {Abbott}
  \emph {et~al.}}]{DES:2022ccp}%
  \BibitemOpen
  \bibfield  {author} {\bibinfo {author} {\bibfnamefont {T.~M.~C.}\
  \bibnamefont {Abbott}} \emph {et~al.} (\bibinfo {collaboration} {DES}),\
  }\href {\doibase 10.1103/PhysRevD.107.083504} {\bibfield  {journal} {\bibinfo
   {journal} {Phys. Rev. D}\ }\textbf {\bibinfo {volume} {107}},\ \bibinfo
  {pages} {083504} (\bibinfo {year} {2023})},\ \Eprint
  {http://arxiv.org/abs/2207.05766} {arXiv:2207.05766 [astro-ph.CO]}
  \BibitemShut {NoStop}%
\bibitem [{\citenamefont {Caldwell}\ \emph {et~al.}(2007)\citenamefont
  {Caldwell}, \citenamefont {Cooray},\ and\ \citenamefont
  {Melchiorri}}]{Caldwell:2007cw}%
  \BibitemOpen
  \bibfield  {author} {\bibinfo {author} {\bibfnamefont {R.}~\bibnamefont
  {Caldwell}}, \bibinfo {author} {\bibfnamefont {A.}~\bibnamefont {Cooray}}, \
  and\ \bibinfo {author} {\bibfnamefont {A.}~\bibnamefont {Melchiorri}},\
  }\href {\doibase 10.1103/PhysRevD.76.023507} {\bibfield  {journal} {\bibinfo
  {journal} {Phys. Rev. D}\ }\textbf {\bibinfo {volume} {76}},\ \bibinfo
  {pages} {023507} (\bibinfo {year} {2007})},\ \Eprint
  {http://arxiv.org/abs/astro-ph/0703375} {arXiv:astro-ph/0703375} \BibitemShut
  {NoStop}%
\bibitem [{\citenamefont {Daniel}\ \emph {et~al.}(2008)\citenamefont {Daniel},
  \citenamefont {Caldwell}, \citenamefont {Cooray},\ and\ \citenamefont
  {Melchiorri}}]{Daniel:2008et}%
  \BibitemOpen
  \bibfield  {author} {\bibinfo {author} {\bibfnamefont {S.~F.}\ \bibnamefont
  {Daniel}}, \bibinfo {author} {\bibfnamefont {R.~R.}\ \bibnamefont
  {Caldwell}}, \bibinfo {author} {\bibfnamefont {A.}~\bibnamefont {Cooray}}, \
  and\ \bibinfo {author} {\bibfnamefont {A.}~\bibnamefont {Melchiorri}},\
  }\href {\doibase 10.1103/PhysRevD.77.103513} {\bibfield  {journal} {\bibinfo
  {journal} {Phys. Rev. D}\ }\textbf {\bibinfo {volume} {77}},\ \bibinfo
  {pages} {103513} (\bibinfo {year} {2008})},\ \Eprint
  {http://arxiv.org/abs/0802.1068} {arXiv:0802.1068 [astro-ph]} \BibitemShut
  {NoStop}%
\bibitem [{\citenamefont {Saltas}\ \emph {et~al.}(2014)\citenamefont {Saltas},
  \citenamefont {Sawicki}, \citenamefont {Amendola},\ and\ \citenamefont
  {Kunz}}]{Saltas:2014dha}%
  \BibitemOpen
  \bibfield  {author} {\bibinfo {author} {\bibfnamefont {I.~D.}\ \bibnamefont
  {Saltas}}, \bibinfo {author} {\bibfnamefont {I.}~\bibnamefont {Sawicki}},
  \bibinfo {author} {\bibfnamefont {L.}~\bibnamefont {Amendola}}, \ and\
  \bibinfo {author} {\bibfnamefont {M.}~\bibnamefont {Kunz}},\ }\href {\doibase
  10.1103/PhysRevLett.113.191101} {\bibfield  {journal} {\bibinfo  {journal}
  {Phys. Rev. Lett.}\ }\textbf {\bibinfo {volume} {113}},\ \bibinfo {pages}
  {191101} (\bibinfo {year} {2014})},\ \Eprint {http://arxiv.org/abs/1406.7139}
  {arXiv:1406.7139 [astro-ph.CO]} \BibitemShut {NoStop}%
\bibitem [{\citenamefont {Sawicki}\ \emph {et~al.}(2017)\citenamefont
  {Sawicki}, \citenamefont {Saltas}, \citenamefont {Motta}, \citenamefont
  {Amendola},\ and\ \citenamefont {Kunz}}]{Sawicki:2016klv}%
  \BibitemOpen
  \bibfield  {author} {\bibinfo {author} {\bibfnamefont {I.}~\bibnamefont
  {Sawicki}}, \bibinfo {author} {\bibfnamefont {I.~D.}\ \bibnamefont {Saltas}},
  \bibinfo {author} {\bibfnamefont {M.}~\bibnamefont {Motta}}, \bibinfo
  {author} {\bibfnamefont {L.}~\bibnamefont {Amendola}}, \ and\ \bibinfo
  {author} {\bibfnamefont {M.}~\bibnamefont {Kunz}},\ }\href {\doibase
  10.1103/PhysRevD.95.083520} {\bibfield  {journal} {\bibinfo  {journal} {Phys.
  Rev.}\ }\textbf {\bibinfo {volume} {D95}},\ \bibinfo {pages} {083520}
  (\bibinfo {year} {2017})},\ \Eprint {http://arxiv.org/abs/1612.02002}
  {arXiv:1612.02002 [astro-ph.CO]} \BibitemShut {NoStop}%
\bibitem [{\citenamefont {Aghanim}\ \emph {et~al.}(2018)\citenamefont {Aghanim}
  \emph {et~al.}}]{Aghanim:2018eyx}%
  \BibitemOpen
  \bibfield  {author} {\bibinfo {author} {\bibfnamefont {N.}~\bibnamefont
  {Aghanim}} \emph {et~al.} (\bibinfo {collaboration} {Planck}),\ }\href@noop
  {} {\  (\bibinfo {year} {2018})},\ \Eprint {http://arxiv.org/abs/1807.06209}
  {arXiv:1807.06209 [astro-ph.CO]} \BibitemShut {NoStop}%
\bibitem [{\citenamefont {Torrado}\ and\ \citenamefont
  {Lewis}(2021)}]{Torrado:2020dgo}%
  \BibitemOpen
  \bibfield  {author} {\bibinfo {author} {\bibfnamefont {J.}~\bibnamefont
  {Torrado}}\ and\ \bibinfo {author} {\bibfnamefont {A.}~\bibnamefont
  {Lewis}},\ }\href {\doibase 10.1088/1475-7516/2021/05/057} {\bibfield
  {journal} {\bibinfo  {journal} {JCAP}\ }\textbf {\bibinfo {volume} {05}},\
  \bibinfo {pages} {057} (\bibinfo {year} {2021})},\ \Eprint
  {http://arxiv.org/abs/2005.05290} {arXiv:2005.05290 [astro-ph.IM]}
  \BibitemShut {NoStop}%
\bibitem [{\citenamefont {Abbott}\ \emph
  {et~al.}(2017{\natexlab{a}})\citenamefont {Abbott} \emph
  {et~al.}}]{TheLIGOScientific:2017qsa}%
  \BibitemOpen
  \bibfield  {author} {\bibinfo {author} {\bibfnamefont {B.}~\bibnamefont
  {Abbott}} \emph {et~al.} (\bibinfo {collaboration} {Virgo, LIGO
  Scientific}),\ }\href {\doibase 10.1103/PhysRevLett.119.161101} {\bibfield
  {journal} {\bibinfo  {journal} {Phys. Rev. Lett.}\ }\textbf {\bibinfo
  {volume} {119}},\ \bibinfo {pages} {161101} (\bibinfo {year}
  {2017}{\natexlab{a}})},\ \Eprint {http://arxiv.org/abs/1710.05832}
  {arXiv:1710.05832 [gr-qc]} \BibitemShut {NoStop}%
\bibitem [{\citenamefont {Abbott}\ \emph
  {et~al.}(2017{\natexlab{b}})\citenamefont {Abbott} \emph
  {et~al.}}]{Monitor:2017mdv}%
  \BibitemOpen
  \bibfield  {author} {\bibinfo {author} {\bibfnamefont {B.~P.}\ \bibnamefont
  {Abbott}} \emph {et~al.} (\bibinfo {collaboration} {Virgo, Fermi-GBM,
  INTEGRAL, LIGO Scientific}),\ }\href {\doibase 10.3847/2041-8213/aa920c}
  {\bibfield  {journal} {\bibinfo  {journal} {Astrophys. J.}\ }\textbf
  {\bibinfo {volume} {848}},\ \bibinfo {pages} {L13} (\bibinfo {year}
  {2017}{\natexlab{b}})},\ \Eprint {http://arxiv.org/abs/1710.05834}
  {arXiv:1710.05834 [astro-ph.HE]} \BibitemShut {NoStop}%
\bibitem [{\citenamefont {Coulter}\ \emph {et~al.}(2017)\citenamefont {Coulter}
  \emph {et~al.}}]{Coulter:2017wya}%
  \BibitemOpen
  \bibfield  {author} {\bibinfo {author} {\bibfnamefont {D.~A.}\ \bibnamefont
  {Coulter}} \emph {et~al.},\ }\href {\doibase 10.1126/science.aap9811}
  {\bibfield  {journal} {\bibinfo  {journal} {Science}\ }\textbf {\bibinfo
  {volume} {358}},\ \bibinfo {pages} {1556} (\bibinfo {year} {2017})},\ \Eprint
  {http://arxiv.org/abs/1710.05452} {arXiv:1710.05452 [astro-ph.HE]}
  \BibitemShut {NoStop}%
\bibitem [{\citenamefont {Baker}\ \emph {et~al.}(2017)\citenamefont {Baker},
  \citenamefont {Bellini}, \citenamefont {Ferreira}, \citenamefont {Lagos},
  \citenamefont {Noller},\ and\ \citenamefont {Sawicki}}]{Baker:2017hug}%
  \BibitemOpen
  \bibfield  {author} {\bibinfo {author} {\bibfnamefont {T.}~\bibnamefont
  {Baker}}, \bibinfo {author} {\bibfnamefont {E.}~\bibnamefont {Bellini}},
  \bibinfo {author} {\bibfnamefont {P.~G.}\ \bibnamefont {Ferreira}}, \bibinfo
  {author} {\bibfnamefont {M.}~\bibnamefont {Lagos}}, \bibinfo {author}
  {\bibfnamefont {J.}~\bibnamefont {Noller}}, \ and\ \bibinfo {author}
  {\bibfnamefont {I.}~\bibnamefont {Sawicki}},\ }\href {\doibase
  10.1103/PhysRevLett.119.251301} {\bibfield  {journal} {\bibinfo  {journal}
  {Phys. Rev. Lett.}\ }\textbf {\bibinfo {volume} {119}},\ \bibinfo {pages}
  {251301} (\bibinfo {year} {2017})},\ \Eprint
  {http://arxiv.org/abs/1710.06394} {arXiv:1710.06394 [astro-ph.CO]}
  \BibitemShut {NoStop}%
\bibitem [{\citenamefont {Ezquiaga}\ and\ \citenamefont
  {Zumalacarregui}(2017)}]{Ezquiaga:2017ekz}%
  \BibitemOpen
  \bibfield  {author} {\bibinfo {author} {\bibfnamefont {J.~M.}\ \bibnamefont
  {Ezquiaga}}\ and\ \bibinfo {author} {\bibfnamefont {M.}~\bibnamefont
  {Zumalacarregui}},\ }\href {\doibase 10.1103/PhysRevLett.119.251304}
  {\bibfield  {journal} {\bibinfo  {journal} {Phys. Rev. Lett.}\ }\textbf
  {\bibinfo {volume} {119}},\ \bibinfo {pages} {251304} (\bibinfo {year}
  {2017})},\ \Eprint {http://arxiv.org/abs/1710.05901} {arXiv:1710.05901
  [astro-ph.CO]} \BibitemShut {NoStop}%
\bibitem [{\citenamefont {Creminelli}\ and\ \citenamefont
  {Vernizzi}(2017)}]{Creminelli:2017sry}%
  \BibitemOpen
  \bibfield  {author} {\bibinfo {author} {\bibfnamefont {P.}~\bibnamefont
  {Creminelli}}\ and\ \bibinfo {author} {\bibfnamefont {F.}~\bibnamefont
  {Vernizzi}},\ }\href {\doibase 10.1103/PhysRevLett.119.251302} {\bibfield
  {journal} {\bibinfo  {journal} {Phys. Rev. Lett.}\ }\textbf {\bibinfo
  {volume} {119}},\ \bibinfo {pages} {251302} (\bibinfo {year} {2017})},\
  \Eprint {http://arxiv.org/abs/1710.05877} {arXiv:1710.05877 [astro-ph.CO]}
  \BibitemShut {NoStop}%
\bibitem [{\citenamefont {Scolnic}\ \emph {et~al.}(2018)\citenamefont {Scolnic}
  \emph {et~al.}}]{Pan-STARRS1:2017jku}%
  \BibitemOpen
  \bibfield  {author} {\bibinfo {author} {\bibfnamefont {D.~M.}\ \bibnamefont
  {Scolnic}} \emph {et~al.} (\bibinfo {collaboration} {Pan-STARRS1}),\ }\href
  {\doibase 10.3847/1538-4357/aab9bb} {\bibfield  {journal} {\bibinfo
  {journal} {Astrophys. J.}\ }\textbf {\bibinfo {volume} {859}},\ \bibinfo
  {pages} {101} (\bibinfo {year} {2018})},\ \Eprint
  {http://arxiv.org/abs/1710.00845} {arXiv:1710.00845 [astro-ph.CO]}
  \BibitemShut {NoStop}%
\bibitem [{\citenamefont {Pogosian}\ and\ \citenamefont
  {Silvestri}(2016)}]{Pogosian:2016pwr}%
  \BibitemOpen
  \bibfield  {author} {\bibinfo {author} {\bibfnamefont {L.}~\bibnamefont
  {Pogosian}}\ and\ \bibinfo {author} {\bibfnamefont {A.}~\bibnamefont
  {Silvestri}},\ }\href {\doibase 10.1103/PhysRevD.94.104014} {\bibfield
  {journal} {\bibinfo  {journal} {Phys. Rev.}\ }\textbf {\bibinfo {volume}
  {D94}},\ \bibinfo {pages} {104014} (\bibinfo {year} {2016})},\ \Eprint
  {http://arxiv.org/abs/1606.05339} {arXiv:1606.05339 [astro-ph.CO]}
  \BibitemShut {NoStop}%
\bibitem [{\citenamefont {Bellini}\ and\ \citenamefont
  {Sawicki}(2014)}]{Bellini:2014fua}%
  \BibitemOpen
  \bibfield  {author} {\bibinfo {author} {\bibfnamefont {E.}~\bibnamefont
  {Bellini}}\ and\ \bibinfo {author} {\bibfnamefont {I.}~\bibnamefont
  {Sawicki}},\ }\href {\doibase 10.1088/1475-7516/2014/07/050} {\bibfield
  {journal} {\bibinfo  {journal} {JCAP}\ }\textbf {\bibinfo {volume} {1407}},\
  \bibinfo {pages} {050} (\bibinfo {year} {2014})},\ \Eprint
  {http://arxiv.org/abs/1404.3713} {arXiv:1404.3713 [astro-ph.CO]} \BibitemShut
  {NoStop}%
\bibitem [{\citenamefont {Aghanim}\ \emph
  {et~al.}(2020{\natexlab{b}})\citenamefont {Aghanim} \emph
  {et~al.}}]{Planck:2018lbu}%
  \BibitemOpen
  \bibfield  {author} {\bibinfo {author} {\bibfnamefont {N.}~\bibnamefont
  {Aghanim}} \emph {et~al.} (\bibinfo {collaboration} {Planck}),\ }\href
  {\doibase 10.1051/0004-6361/201833886} {\bibfield  {journal} {\bibinfo
  {journal} {Astron. Astrophys.}\ }\textbf {\bibinfo {volume} {641}},\ \bibinfo
  {pages} {A8} (\bibinfo {year} {2020}{\natexlab{b}})},\ \Eprint
  {http://arxiv.org/abs/1807.06210} {arXiv:1807.06210 [astro-ph.CO]}
  \BibitemShut {NoStop}%
\bibitem [{\citenamefont {Beutler}\ \emph {et~al.}(2011)\citenamefont
  {Beutler}, \citenamefont {Blake}, \citenamefont {Colless}, \citenamefont
  {Jones}, \citenamefont {Staveley-Smith}, \citenamefont {Campbell},
  \citenamefont {Parker}, \citenamefont {Saunders},\ and\ \citenamefont
  {Watson}}]{Beutler:2011hx}%
  \BibitemOpen
  \bibfield  {author} {\bibinfo {author} {\bibfnamefont {F.}~\bibnamefont
  {Beutler}}, \bibinfo {author} {\bibfnamefont {C.}~\bibnamefont {Blake}},
  \bibinfo {author} {\bibfnamefont {M.}~\bibnamefont {Colless}}, \bibinfo
  {author} {\bibfnamefont {D.~H.}\ \bibnamefont {Jones}}, \bibinfo {author}
  {\bibfnamefont {L.}~\bibnamefont {Staveley-Smith}}, \bibinfo {author}
  {\bibfnamefont {L.}~\bibnamefont {Campbell}}, \bibinfo {author}
  {\bibfnamefont {Q.}~\bibnamefont {Parker}}, \bibinfo {author} {\bibfnamefont
  {W.}~\bibnamefont {Saunders}}, \ and\ \bibinfo {author} {\bibfnamefont
  {F.}~\bibnamefont {Watson}},\ }\href {\doibase
  10.1111/j.1365-2966.2011.19250.x} {\bibfield  {journal} {\bibinfo  {journal}
  {Mon. Not. Roy. Astron. Soc.}\ }\textbf {\bibinfo {volume} {416}},\ \bibinfo
  {pages} {3017} (\bibinfo {year} {2011})},\ \Eprint
  {http://arxiv.org/abs/1106.3366} {arXiv:1106.3366 [astro-ph.CO]} \BibitemShut
  {NoStop}%
\bibitem [{\citenamefont {Ross}\ \emph {et~al.}(2015)\citenamefont {Ross},
  \citenamefont {Samushia}, \citenamefont {Howlett}, \citenamefont {Percival},
  \citenamefont {Burden},\ and\ \citenamefont {Manera}}]{Ross:2014qpa}%
  \BibitemOpen
  \bibfield  {author} {\bibinfo {author} {\bibfnamefont {A.~J.}\ \bibnamefont
  {Ross}}, \bibinfo {author} {\bibfnamefont {L.}~\bibnamefont {Samushia}},
  \bibinfo {author} {\bibfnamefont {C.}~\bibnamefont {Howlett}}, \bibinfo
  {author} {\bibfnamefont {W.~J.}\ \bibnamefont {Percival}}, \bibinfo {author}
  {\bibfnamefont {A.}~\bibnamefont {Burden}}, \ and\ \bibinfo {author}
  {\bibfnamefont {M.}~\bibnamefont {Manera}},\ }\href {\doibase
  10.1093/mnras/stv154} {\bibfield  {journal} {\bibinfo  {journal} {Mon. Not.
  Roy. Astron. Soc.}\ }\textbf {\bibinfo {volume} {449}},\ \bibinfo {pages}
  {835} (\bibinfo {year} {2015})},\ \Eprint {http://arxiv.org/abs/1409.3242}
  {arXiv:1409.3242 [astro-ph.CO]} \BibitemShut {NoStop}%
\bibitem [{\citenamefont {Alam}\ \emph {et~al.}(2017)\citenamefont {Alam} \emph
  {et~al.}}]{BOSS:2016wmc}%
  \BibitemOpen
  \bibfield  {author} {\bibinfo {author} {\bibfnamefont {S.}~\bibnamefont
  {Alam}} \emph {et~al.} (\bibinfo {collaboration} {BOSS}),\ }\href {\doibase
  10.1093/mnras/stx721} {\bibfield  {journal} {\bibinfo  {journal} {Mon. Not.
  Roy. Astron. Soc.}\ }\textbf {\bibinfo {volume} {470}},\ \bibinfo {pages}
  {2617} (\bibinfo {year} {2017})},\ \Eprint {http://arxiv.org/abs/1607.03155}
  {arXiv:1607.03155 [astro-ph.CO]} \BibitemShut {NoStop}%
\end{thebibliography}%

\end{document}